\documentclass{article}
\usepackage{jheppub,esint,shuffle,psfrag}
 \usepackage[utf8]{inputenc}
\usepackage{xcolor}
\usepackage{varioref}
\usepackage{amsmath,amsfonts,amsthm,mathrsfs}
\usepackage{enumerate}
\usepackage{float}
\usepackage{fancyvrb}
\usepackage{verbatim}
\usepackage{wrapfig}
\usepackage{appendix}
\usepackage{amstext}
\usepackage{amssymb}
\usepackage{graphicx}
\usepackage{color}
\usepackage{varioref}
\usepackage{multirow,graphics}
\usepackage{epstopdf}

\usepackage{bbm}
\usepackage{slashed}
\usepackage{relsize}

\newcommand{\com}[1]{(*{\textbf{#1}}*)}

\newcommand{\enrico}[1]{\com{E: #1}}

\numberwithin{equation}{section}

\usepackage{tikz}
\usetikzlibrary{plotmarks,calc,decorations, decorations.pathmorphing, patterns}
\usetikzlibrary{arrows}
\tikzstyle dynkin node=[very thick,shape=circle,draw,inner sep=0pt,minimum size=5mm]
\tikzstyle dynkin line=[very thick]
\tikzstyle inverse line=[gray,line width=1.46pt,line cap=round, dash pattern=on 0pt off 2\pgflinewidth]
\tikzstyle red phase=[red,decoration={snake,amplitude=0.1mm,segment length=1.6mm},decorate]
\tikzstyle blue phase=[blue,decoration={snake,amplitude=0.1mm,segment length=0.9mm},decorate]
\tikzstyle green phase=[green,decoration={snake,amplitude=0.1mm,segment length=0.9mm},decorate]
\tikzstyle brown phase=[brown,decoration={snake,amplitude=0.1mm,segment length=0.9mm},decorate]
\newcommand{\boundellipse}[3]% center, xdim, ydim
{(#1) ellipse (#2 and #3)
}
\usetikzlibrary{decorations.pathmorphing}
\tikzstyle arrow=[thick,rounded corners=18pt,-latex]
\tikzstyle box=[draw,rounded corners,outer sep=4pt]
\tikzstyle B node=[outer sep=0pt]
\tikzstyle Q node=[inner sep=1pt,outer sep=0pt]
\definecolor{purple_nice}{rgb}{0.4,0.2,0.7}
\definecolor{fuel_blue}{RGB}{42,162,185}

\def\<{\langle}
\def\>{\rangle}

\newcommand{\p}{\partial}

\makeatletter
\@ifundefined{usebibtex}{} {}
\makeatother

\def\Tr{\text{Tr}~}

\title{
\Large Conformal quantum mechanics \& the integrable spinning Fishnet}
\author{Sergey Derkachov$^a$ and Enrico Olivucci$^b$}
\affiliation[a]{St. Petersburg Department of the Steklov Mathematical Institute
of Russian Academy of Sciences,
Fontanka 27, 191023 St. Petersburg, Russia.}
\affiliation[b]{Perimeter Institute for Theoretical Physics, Waterloo, Ontario N2L2Y5, Canada.}

\usepackage{esint}
\usepackage{breqn}
\def \Tr {\mathop{\rm Tr}\nolimits}

\newcommand{\sig}{\boldsymbol{\sigma}}
\newcommand{\bsig}{\boldsymbol{\bar \sigma}}
\newcommand{\rmat}{\mathcal{R}}

\def\numberbysection{\@addtoreset{equation}{section}
                     \def\theequation{\thesection.\arabic{equation}}}

\begin{document}
\abstract{In this paper we consider systems of quantum particles in the $4d$ Euclidean space which enjoy conformal symmetry. The algebraic relations for conformal-invariant combinations of positions and momenta are used to construct a solution of the Yang-Baxter equation in the unitary irreducibile representations of the principal series $\Delta=2+i\nu$ for any left/right spins $\ell,\dot{\ell}$ of the particles.  Such relations are interpreted in the language of Feynman diagrams as integral \emph{star-triangle} identites between propagators of a conformal field theory.  We prove the quantum integrability of a spin chain whose $k$-th site hosts a particle in the representation $(\Delta_k,\ell_k,
\dot{
\ell}_k)$ of the conformal group,  realizing a spinning and inhomogeneous version of the quantum magnet used to describe the spectrum of the bi-scalar Fishnet theories \cite{Gurdogan:2015csr}. For the special choice of particles in the scalar $(1,0,0)$ and fermionic $(3/2,1,0)$ representation the transfer matrices of the model are Bethe-Salpeter kernels for the double-scaling limit  of specific two-point correlators in the  $\gamma$-deformed $\mathcal{N}=4$ and $\mathcal{N}=2$ supersymmetric theories.}
\maketitle
\section{Introduction}
%The study of exactly-solvable models of quantum particles with conformal symmetry lies at the intersection between integrability and conformal field theory (CFT), which has attracted a major interest by theoretical physicists in the last decades. The search for a quantum description of gravity by means of string theory - a $(1+1)$-dimensional conformal field theory - and the remarkable results delivered by conformal bootstrap techniques in the theory of second order phase transitions are among the main reason that justify the large attention devoted to conformal symmetry. A remarkable example of a field theory which enjoys conformal symmetry and is solvable exactly, i.e. the scaling dimension of operators can be computed at finite coupling, is given by the $\mathcal{N}=4$ super-Yang-Mills theory. The spectral problem for the operators of the theory has been solved starting from a one-loop observation: the computation of Feynman integrals appearing at this order in two point functions can be mapped to the diagonalization of the Hamiltonian of an appropriate generalization of the integrable Heisenberg quantum magnet i.e. to the solution of a set of finite-difference Baxter (or Bethe) equations \cite{}. By the assumption of integrability at finite coupling, the Baxter equation can be lifted to a Quantum Spectral Curve \cite{}, which can be computed explicitly, and higher loop checks have confirmed the guess of integrability despite a direct derivation of it, and the absence of a concrete spin-chain or other associated quantum system.

This work is based on the papers~\cite{Isaev_2003,Chicherin:2012yn,Derkachov2020,Derkachov:2020zvv}; our motivation is to clarify their various interrelations and to deduce how their results apply to the most general setup. First of all we managed to reformulate the expression given in~\cite{Chicherin:2012yn} for the intertwiners $S_1,S_2,S_3$ of unitary irreducible representations of $SO(1,5)$, and thus the infinite-dimensional R-operator, in the form of symmetric traceless tensors of coordinates and momenta~\cite{Isaev_2003,Isaev:2007uy}. Ultimately, these achievements are based on the new star-triangle relation derived in~\cite{Derkachov2020}. On the operative level we developed an handful \emph{Feynman diagram} notation of lines and dots where the algebraic calculations based on the star-triangle duality are realized as graphical operations, providing ground for an extension to spinning conformal fields of the techniques~\cite{Vasiliev:1982dc,Vasiliev:1981dg,Kazakov:1983ns,Kazakov:1983pk,Kazakov:1984km} (see also the software implementation \cite{Preti:2018vog,Preti_2020}).
Secondly, by the knowledge of the R-operator we can formulate a new quantum integrable model with conformal symmetry as a chain of particles in the space $\mathbb{R}^4$, each of them transforming in a given representations of the principal series~\cite{Tod:1977harm}. The transfer matrices of the model are identified with \emph{graph-building} operators of a planar Feynman integral with square-lattice topology and spinning propagators.

The organization of the paper is the following.
In the section \ref{cqm_str} we present a compact formula for the Fourier
transformation of the general conformal propagator in the representation of scaling dimension $\Delta$ and spins $\ell/2$ and $\dot{\ell}/2$. Along with it we introduce the generalized star-triangle duality involving such propagators and we explain the diagrammatic notations.
The duality allows to derive by few graphical steps two main interchange
relations which are ubiquitous in the proofs of commutativity of
the transfer matrices of the integrable chain.
Section~\ref{sect:gen} is devoted to the construction of the
R-operator for general principal series representations, and the proof of the corresponding Yang-Baxter relation is delivered in diagrammatic form. As a whole, the results of the first two sections contain the four-dimensional version of the $SL(2,\mathbb{C})$ techniques developed in \cite{Derkachov:2001yn} for $2d$ conformal quantum models.

In section~\ref{InFish} we construct the integrable chain starting from its transfer matrices. We propose a factorization of the transfer matrices in the product of
two commuting operators and use it to demonstrate the quantum integrability of the model. For specific choices of the representations of the chain's particles the integral kernel of the transfer matrix operator is the Bethe-Salpeter kernel for the planar two-point functions of various Fishnet/chiral conformal field theories \cite{Gurdogan:2015csr,Caetano:2016ydc,Kazakov_2018,Kazakov_2019,Pittelli2019}.

\section{Conformal quantum mechanics and star-triangle relation}
\label{cqm_str}
Let us consider a quantum particle with left/right spins $\frac{\ell}{2}$ and $\frac{\dot{\ell}}{2}$ moving in the $4d$ Euclidean space. Its state is described by a wave function of the position $x=(x_0,x_1,x_2,x_3)$ with $\ell$- and $\dot{\ell}$- symmetric spinor indices
\begin{align}
\begin{aligned}
\label{wave_f}
 &\Phi_{\mathbf{a\dot{a}}}(x)=\Phi_{(a_1\dots a_{\ell})(\dot{a}_1\dots\dot{a}_{\dot{\ell}})}(x)\,,\,\,\,\,\, a_i,\dot{a}_i
\in \{1,2\}\,,
\end{aligned}
\end{align}
where $(a_1\dots a_{\ell})$ means symmetrization
$(\dots a\dots b\dots )=(\dots b\dots a\dots )$.

The  wavefunctions \eqref{wave_f}  belong to the Hilbert space
\begin{equation}
\label{Hilbert}
\mathbb{V}\simeq L^2(d^4x)\otimes \text{Sym}_{\ell}[\mathbb{C}^2] \otimes \text{Sym}_{\dot{\ell}}[\mathbb{C}^2]\,,
\end{equation}
with $\text{Sym}_{
\ell}[\mathbb{C}^2]\subset (\mathbb{C}^{2})^{\otimes \ell} $ the space of complex symmetric spinors $u_{\mathbf{a}}=u_{(a_1,\dots,a_{\ell})}$.  The scalar product on \eqref{Hilbert} is inherited from the standard ones defined on its factors, that is
\begin{equation}
\label{scalar_product}
\langle F , G\rangle_{\mathbb{V}} \,=\,  \int d^4 x\, \langle F(x) |G(x) \rangle\,=\,\int d^4 x\, (F^*)^{\mathbf{a}\mathbf{\dot{a}}}(x) \,G_{\mathbf{a}\mathbf{\dot{a}}}(x)\,.
\end{equation}
The spinors in paper have lower indices $\alpha_{a}$ and $\beta_{\dot{a}}$
and the index raising operation is defined as complex conjugation $\alpha^{a} =(\alpha_{a})^*=\bar{\alpha}_{a}$ and
$\beta^{\dot{a}} =(\beta_{\dot{a}})^*=\bar{\beta}_{\dot{a}}$ and the pairing
between spinors is the standard scalar product in $\mathbb{C}^2$
\begin{align}
\langle\alpha|\alpha'\rangle =
\alpha^{a} \alpha'_{a} = \bar{\alpha}_{a} \alpha'_{a}\, ,\, \,\,\,
\langle\beta'|\beta\rangle = {\beta'}^{\dot{a}}\beta_{\dot{a}} =
\bar{\beta}'_{\dot{a}}\beta_{\dot{a}}\,.
\end{align}

A rotation of the Euclidean space $x^{
\mu} \mapsto \Lambda_{\nu}^{\mu}x^{\nu}$ is represented on the symmetric spinors according to the decomposition $SO(4)\simeq SU(2)\times SU(2)$, by two matrices $U,V$ rotating the dotted/undotted  indices
\begin{align}
\label{spinor_rotation}
\begin{aligned}
&u_{\mathbf{a}} \mapsto u_{\mathbf{a}} '=[U]^{\mathbf{b}}_{\mathbf{a}} u_{\mathbf{b}}\,,\,\,\,\,\,\,\,\, v_{\mathbf{\dot a}} \mapsto v_{\mathbf{\dot a}} '=[V]^{\mathbf{\dot b}}_{\mathbf{\dot a}} v_{\mathbf{\dot b}}\,,
\end{aligned}
\end{align}
and the following notations will be frequently used throughout the text
\begin{align}
\begin{aligned}
&\mathbf{a}\equiv (a_1,\dots,a_{\ell})\,,\,\,\,\mathbf{\dot a}\equiv (\dot  a_1,\dots,\dot a_{\dot \ell})\,,\,\,\,a_k,\dot{a}_k \in \{1,2\}\,,\\
&u_{\mathbf{a}} = u_{(a_1\dots a_{\ell})}\,,\,\,v_{\mathbf{\dot a}} = v_{(\dot a_1\dots \dot a_{\ell})}\,,\,\,\,\,[U]_{\mathbf{a}}^{\mathbf{b}} = U_{(a_1}^{(b_1}\cdots U_{a_{\ell})}^{b_{\ell})}\,.
\end{aligned}
\end{align}
Alternatively we will use the compact notation $[U]^{\ell}$ which specifies the spin number $\ell$ without explicit spinor indices
\begin{equation}
\label{compact_xp}
([U]^{\ell})_{\mathbf{a}}^{\mathbf{b}} = U_{(a_1}^{(b_1}\cdots U_{a_{\ell})}^{b_{\ell})}\,.
\end{equation}
We are interested in quantum systems with conformal symmetry,  that is in particles evolving under the action of an Hamiltonian operator $\mathcal{H}$ which is invariant under conformal transformation
\begin{equation}
x^{\mu}\mapsto y^{\mu}(x)\,,\,\,\,\,\,\,\,\frac{\partial y^{\mu}(x)}{\partial x^{\kappa}}\frac{\partial y_{\nu}(x)}{\partial x_{\kappa}}= \Lambda(y)\,\delta^{\mu}_{\nu}\,.
\end{equation}
Examples of such systems in a $d=2$ dimensional spacetime are the $SL(2,\mathbb{C})$ integrable spin chains introduced by Lipatov to describe scattering amplitudes of QCD gluons at high energy~\cite{Lipatov:1993qn,Lipatov:1993yb,Faddeev:1994zg}. Recently, higher dimensional ($d>2$) versions of conformal spin chains have attracted attention in the realm of Fishnet conformal field theories \cite{Gurdogan:2015csr,Kazakov:2018qez,Caetano:2016ydc,Gromov:2017cja,Derkachov2019,
BassoFerrando,Derkachov2020}, as they describe the Feynman integrals of the theory at any coupling and therefore the anomalous dimensions of operators at finite coupling.  In a conformal invariant system the eigenfunctions of the model should transform under an irreducible unitary representation of the conformal group.  Hence, we study wavefunctions transforming as
\begin{equation}
\label{princ_rep}
x^{\mu}\mapsto y^{\mu}(x)\,,\,\,\,\,\,\,\,\Phi_{\mathbf{a\dot{a}}}(x)\,\mapsto\, {\Phi'}_{\mathbf{a\dot{a}}}(x)\,=\Lambda(y)^{\Delta}\,[U]_{\mathbf{a}}^{\,\mathbf{b}} [V]^{\,\mathbf{\dot{b}}}_{\mathbf{\dot{a}}} \,\Phi_{\mathbf{b \dot{b}}}(y) \,,
\end{equation}
where $U_{a}^b$ and $V_{\dot{a}}^{\dot{b}}$ are two $SU(2)$ matrices that realize the spinor representation of $SO(4)$ rotations. The dotted and un-dotted indices distinguish
spinors that are transformed according to representations
of two different copies of $SU(2)$. 

Without loss of generality we can restrict ourselves to the representations of the principal series,  that means $\Delta=2+i\nu$ with $\nu$ a real number\footnote{The unitary irreducible representations of $SO(1,5)$ - as explained in \cite{Tod:1977harm} - would include also the complementary series,  for which $0<\Delta<4$. Our results throughout the paper can be extended to this class of representations by an analytic continuation of $\nu$ to the imaginary segment $i[-2,2]$.}.  A scale-invariant combination of the position operator $x^{\mu}$ and momentum
operator $\hat p_{\mu}=-i\partial_{\mu}$ satisfies the remarkable duality~\cite{Isaev_2003,Isaev:2007uy}
\begin{equation}
\label{STR_scalar_op}
\hat p^{2u} x^{2(u+v)} \hat p^{2v} = x^{2v} \hat p^{2(u+v)} x^{2u}\,,
\end{equation}
where the operator $\hat p^{2u}$ is understood as an integral operator
\begin{equation}
\label{p_def}
\hat p^{2u} f(x)= \frac{1}{a(u)}\,\int d^4 y \frac{f(y)}{(x-y)^{2(u+2)}}\,.
\end{equation}
Note that we shall consider the generic situation $u \in \mathbb{C}$ and understand
all similar integrals as an analytic continuation in $u$ from the 
convergence domain \cite{Gelfand:105396}.

The representation \eqref{p_def} follows from the formula for the Fourier transformation of a power function
\begin{align}\label{Fscalar}
\int d^4 x \,\frac{e^{ipx}}{x^{2(u+2)}} = a(u)\, p^{2u}\,\,\,,\,\,
a(u) = \frac{\pi^2\,\Gamma(-u)}{4^{u}\Gamma(u+2)}\,,
\end{align}
where $\Gamma(u)$ is the Euler gamma-function.

\begin{figure}[H]
\begin{center}
\includegraphics[scale=1.5]{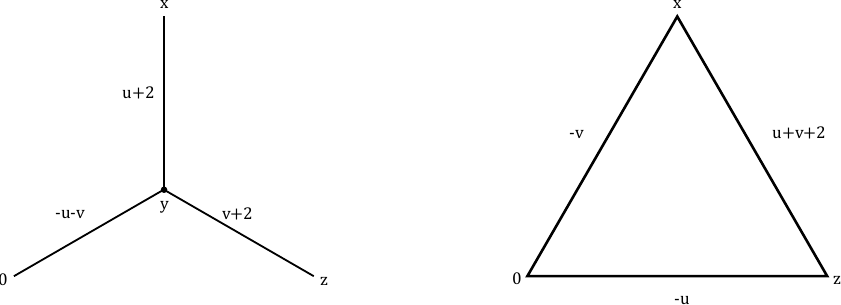}
\caption{\textbf{Left}: Convolution of three scalar propagators corresponding to the l.h.s. of the duality \eqref{STR_scalar_ker}. Each solid segment is 1/$x_{ij}^2$ to the power written next to it, where $x_i$ and $x_j$ are the endpoints of the segment. \textbf{Right}: product of three propagators corresponding to the r.h.s. of \eqref{STR_scalar_ker} modulo a multiplicative pre-factor.}
\label{scalar_STR}
\end{center}
\end{figure}

According to \eqref{p_def} the equation \eqref{STR_scalar_op} has a straightforward interpretation in the language of Feynman diagrams as a \emph{star-triangle} identity as illustrated in Fig.\ref{scalar_STR}.
Indeed, \eqref{STR_scalar_op} is an identity between two integral operators and the application of both sides of the equation
to the delta-function $\delta^{(4)}(x-z)$ delivers an equality between the corresponding kernels
\begin{align}
\label{STR_scalar_ker}
\int  \frac{d^4y}{(x-y)^{2(u+2)}y^{-2(u+v)}(y-z)^{2(v+2)}} =
\frac{a(u)\,a(v)}{a(u+v)}\,\frac{1}{x^{-2v}(x-z)^{2(u+v+2)}z^{-2u}}\,,
\end{align}
that in terms of Feynman diagrams relates a vertex of three propagators whose scaling dimensions sum up to $d=4$ - the \emph{star} integral - with the product of three propagators between the vertices of the star - the \emph{triangle}.
In the following we generalize the equality \eqref{STR_scalar_op} to the case of spinning particles $\ell,\dot{\ell}\neq 0$. Therefore, a scalar propagators $x^{2u}$ of a field of dimension $\Delta=-u$ is replaced by the propagator of a conformal field in the representation $(\Delta=-u,\ell,\dot{\ell})$, that stripped of all constants reads~\cite{Sotkov:1976xe}
\begin{equation}
\label{prop_CFT_1}
G_{\ell\dot{\ell}}(u,x) = x^{2u}\,[\mathbf{\overline{x}}]^{\ell}
[\mathbf{x}]^{\dot{\ell}}\,,\,\,\,\mathbf{x}=\boldsymbol{\sigma}_{\mu}\frac{x^{\mu}}{(x^2)^{\frac{1}{2}}}\,,\,\,\,\,\, \mathbf{\overline x}=\boldsymbol{\overline{\sigma}}_{\mu}\frac{x^{\mu}}{(x^2)^{\frac{1}{2}}}\,.
\end{equation}
Here the matrices $\boldsymbol{
\sigma}_{\mu}$ and $\boldsymbol{
\overline{\sigma}}_{\mu}$ are defined in terms of the standard Pauli matrices as $\boldsymbol{\sigma}_k = i\sigma_k$, $\boldsymbol{\overline{\sigma}}_k = -i\sigma_k$ for $k=1,2,3$ and $\boldsymbol{\sigma}_0=\boldsymbol{\sigma}_0=\mathbbm{1}$,  so that they satisfy the Clifford algebra
\begin{equation}
\boldsymbol{
\sigma}_{\mu}\boldsymbol{
\overline\sigma}_{\nu}+\boldsymbol{
\sigma}_{\nu}\boldsymbol{
\overline\sigma}_{\mu} = 2\delta_{\mu\nu}\,.
\end{equation}
Everywhere in the paper we work with Euclidean metric
and it is not useful for our purposes to distinguish
carefully the upper and lower tensor indices like $\mu\,,\nu$.
For the sake of simplicity for the cumbersome formulae we adopt the
following notation in the paper for spinor indices
$\boldsymbol{\sigma}_{\mu}= ||(\boldsymbol{\sigma}_{\mu})_{a}^{\dot{a}}||$
and $\overline{ \boldsymbol{\sigma} }_{\mu} = ||(\overline{ \boldsymbol{\sigma} }_{\mu})_{\dot{a}}^{a}||$ so that \eqref{prop_CFT_1} in explicit notations
with all spinor indices has the following form
\begin{equation}
\label{propag_CFT}
G_{\ell\dot{\ell}}(u,x)_{\mathbf{\dot{a}a}}^{\mathbf{b\dot{b}}}\,=\,
x^{2u}\,[\mathbf{\overline{x}}]_{\mathbf{\dot{a}}}^{\mathbf{b}}
[\mathbf{x}]_{\mathbf{a}}^{\mathbf{\dot{b}}}\,.
\end{equation}
Of course the same generalization of $\hat p^{2u}\rightarrow {G}_{\ell\dot{\ell}}(u,p)$ can be done for the momentum operator. The main difference w.r.t. the case of the scalar propagators
is that the Fourier transformation of \eqref{propag_CFT} involves a mixing of spinorial indices:
\begin{align}
\label{Fspinor2}
&\int d^4 x e^{ipx}\,\frac{[\mathbf{\overline{x}}]^{\ell}
[\mathbf{x}]^{\dot{\ell}}}{x^{2(u+2)}} = a_{\ell\dot{\ell}}(u)\,
\hat p^{2u}\,[\mathbf{\overline{p}}]^{\ell}\mathbf{R}_{\ell\dot{\ell}}(u)
[\mathbf{p}]^{\dot{\ell}}\,,
\end{align}
where the constant $a_{\ell\dot{\ell}}(u)$ is a generalization of the constant $a(u)$ to general non-zero spins $\ell,
\dot{\ell}$
\begin{align}
a_{\ell\dot{\ell}}(u) =  \frac{(-i)^{\ell+\dot{\ell}}\,\pi^2\,\Gamma\left(-u-\frac{\ell+\dot{\ell}}{2}\right)
\,\Gamma\left(u+\frac{\ell+\dot{\ell}}{2}+1\right)}
{4^{u}\,\left(u+\frac{\ell+\dot{\ell}}{2}+1\right)\,
\Gamma\left(u+1+\frac{\ell-\dot{\ell}}{2}\right)\,
\Gamma\left(u+1+\frac{\dot{\ell}-\ell}{2}\right)}\,.
\end{align}
The matrix $\mathbf{R}_{nm}(u)$ appearing in the r.h.s. of \eqref{Fspinor2} is the compact notation for
\begin{equation}
\label{RR_fused}
\mathbf{R}_{\mathbf{ac}}^{\mathbf{bd}}(u) =
\mathbf{R}_{(a_1\dots a_n)(c_1\dots c_m)}^{(b_1\dots b_n)(d_1\dots d_m)}(u)\,,
\end{equation}
which is a solution of the Yang-Baxter equation (YBE) on the space
of $n$- and $m$-symmetric spinors
\begin{equation}
\mathbf{R}_{nm}(u-v)\mathbf{R}_{n\ell}(u)\mathbf{R}_{m\ell}(u)=
\mathbf{R}_{m\ell}(v)\mathbf{R}_{n\ell}(u)\mathbf{R}_{nm}(u-v)\,.
\end{equation}
It can be obtained by the \emph{fusion} procedure \cite{Kulish:1981gi,Kulish:1981bi} starting from the well-known solution in the defining representation $n=m=1$
\begin{equation}
\mathbf{R}_{ac}^{bd}(u) =\frac{1}{u+1}\left(u\, \delta_{a}^b \delta_c^d+\delta_{a}^d \delta_c^b\right): \mathbb{C}^2\otimes\mathbb{C}^2 \longrightarrow \mathbb{C}^2\otimes \mathbb{C}^2\,.
\end{equation}
and the final expression is given by (C.12)-(C.14) of the paper~\cite{Derkachov:2020zvv}.
Besides the YBE, the $\mathbf R$-matrix satisfies the parity relation $\mathbf{R}_{nm}(u)\mathbf{R}_{nm}(-u) = \mathbbm{1}$, or in explicit form
\begin{equation}
\label{unitar}
\mathbf{R}_{\mathbf{a_1a_2}}^{\mathbf{c_1c_2}}(u)
\mathbf{R}_{\mathbf{c_1c_2}}^{\mathbf{b_1b_2}}(-u)=
\delta_{\mathbf{a_1}}^{\mathbf{b_1}}\delta_{\mathbf{a_2}}^{\mathbf{b_2}}\,,
\end{equation}
and the crossing symmetry $\mathbf{R}^t_{nm}(u)\mathbf{R}^t_{nm}(-u-2) = r_{nm}(u)\,\mathbbm{1}$,
\begin{align}
\label{crossinxg}
&\mathbf{R}_{\mathbf{c_1a_2}}^{\mathbf{a_1c_2}}(u)
\mathbf{R}_{\mathbf{b_1c_2}}^{\mathbf{c_1b_2}}(-u-2) = r_{nm}(u)\,
\delta_{\mathbf{b_1}}^{\mathbf{a_1}}\delta_{\mathbf{a_2}}^{\mathbf{b_2}}
\end{align}
with the scalar factor
\begin{align*}
r_{nm}(u) = \frac{\left(u+\frac{n+m}{2}+1\right)}{\left(u+\frac{n-m}{2}+1\right)}
\frac{\left(-u+\frac{n+m}{2}-1\right)}{\left(-u+\frac{n-m}{2}-1\right)}\,,
\end{align*}
and $t$ means transposition in the space of $n$-symmetric spinors.
The crossing symmetry \eqref{crossinxg} allows to rewrite the equality \eqref{Fspinor2} in the equivalent form
\begin{align}
\label{Fspinor1}
&r_{\ell\dot{\ell}}(u)\,a_{\ell\dot{\ell}}(u)\, \hat p^{2u}\,[\mathbf{\overline{p}}]^{\ell}\,[\mathbf{p}]^{\dot{\ell}}\,=\,\int d^4 x  e^{ipx}\,\frac{[\mathbf{\overline{x}}]^{\ell}
\mathbf{R}_{\ell\dot{\ell}}(-u-2)[\mathbf{x}]^{\dot{\ell}}}{x^{2(u+2)}}\,.
\end{align}

\begin{figure}[H]
\begin{center}
\includegraphics[scale=2]{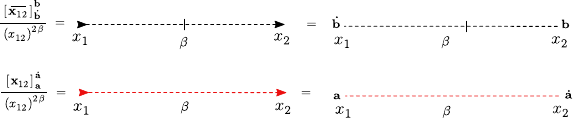}
\caption{Graphic representation of the propagator $G_{\ell\dot{\ell}}(u,x_1-x_2)$ of \eqref{propag_CFT} with conformal dimension $\beta = -u$ and spins $(\ell,\dot{\ell})=(0,\dot \ell)$ - first row - or $(\ell,\dot{\ell})=(\ell,0)$ - second row. The dashed lines stand for the spinorial structures in \eqref{propag_CFT}, the one with the bar denoting the matrix ${\mathbf{\overline x}}$. The arrows denote the flow of spinor indices.}
\label{fig_sing}
\end{center}
\end{figure}

\begin{figure}[H]
\begin{center}
\includegraphics[scale=2.3]{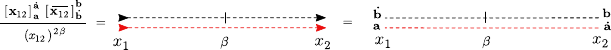}
\caption{Graphic representation of the propagator $G_{\ell\dot{\ell}}(u,x_1-x_2)$ of \eqref{propag_CFT} with conformal dimension $\beta=-u$ and generic spins $\ell,\dot{\ell}$. The two dashed lines stand for the spinorial structures in \eqref{propag_CFT}, the one with the bar denoting the matrix ${\mathbf{\overline x}}$.  The arrows denote the flow of spinor indices, as made explicit by the r.h.s.}
\label{fig_propag}
\end{center}
\end{figure}

In terms of Feynman diagrams we represent the spinning propagators \eqref{propag_CFT} by a double dashed line, see Figg.\ref{fig_sing},\ref{fig_propag}.  Each line stands for the spinor structure $[\mathbf{x}]$ or $[\mathbf{\overline{x}}]$, as in Fig.\ref{fig_sing}, and we use arrows to indicate the flow of spinor indices.
\begin{figure}[H]
\begin{center}
\includegraphics[scale=2]{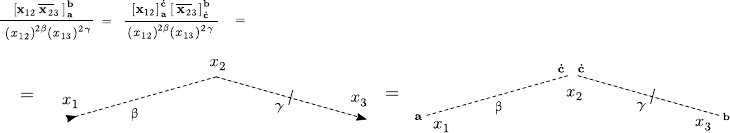}
\caption{Graphic representation of the product of two propagators respect to their spinor structure. The arrows denote the order of the matrices $[\mathbf{x}]$ and $[\overline{\mathbf{x}}]$ in the product.}
\label{fig_conv}
\end{center}
\end{figure}
\begin{figure}[H]
\begin{center}
\includegraphics[scale=1.05]{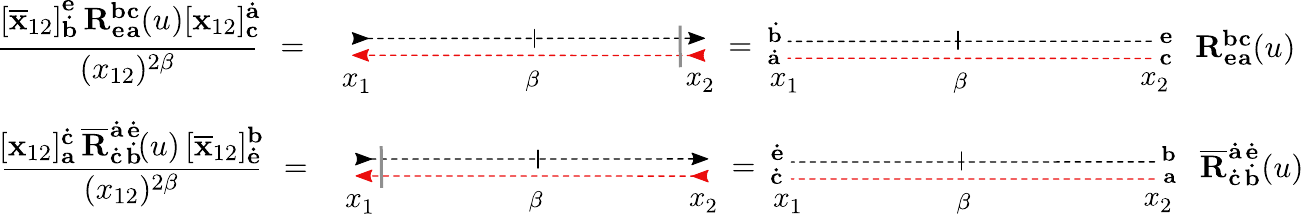}
\caption{Graphic representation of the propagator $S_{\ell\dot{\ell}}(u,x_1-x_2)$ of \eqref{S_def}. The two dashed lines stand for the spinorial structures in \eqref{propag_CFT}, and the mixing matrix $\mathbf{R}$ (or $\overline{\mathbf{R}}$) is inserted along their structure as a solid grey line. On the right we show the explicit spinor structure of $S_{\ell\dot{\ell}}(u,x_1-x_2)$, corresponding to the compact notation on the left.}
\label{fig_propag_R}
\end{center}
\end{figure}
\noindent
In fact, the matrices $\mathbf{R}(u)$ and $\mathbf{\overline{R}}(u)$ differ only for the type of indices (un-dotted/dotted) but their matrix elements are identical, thus we'll often omit to specify the over-lined notation in the following. Moreover, pair of lines of the type appearing in the notation of Fig.\ref{fig_propag} are always assumed to be one of type $[\mathbf{x}] $ and one of type $[\mathbf{\overline x}]$, even when not specified by the graphical notation. Similarly, two convoluted lines as in Fig.\ref{fig_conv} must be always $[\mathbf{x}] [\mathbf{\overline x}]$ either $[\mathbf{\overline x}][\mathbf{x}]$. All the formulas that we write in the paper are actually invariant under the exchange $\boldsymbol{\sigma} \leftrightarrow \boldsymbol{\overline{\sigma}}$.

\begin{figure}[H]
\begin{center}
\includegraphics[scale=1.8]{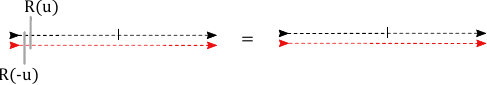}\\
\includegraphics[scale=1.8]{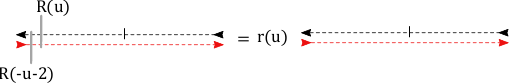}
\end{center}
\caption{\textbf{Up}: Parity property of the matrix $\mathbf{R}_{mn}(u)$. The product of two matrices with opposite arguments along two spinorial structure with the same orientation cancels. \textbf{Down}: Crossing symmetry of the matrix $\mathbf{R}_{mn}(u)$. The two matrices act along two spinor structures flowing in the opposite directions,  according to the arrows,  so their product is transposed in one of the two spinor spaces.}
\label{unit_cross_R}
\end{figure}

 Let us state the rules
\begin{align}
&[\mathbf{x}]_{\mathbf{a}}^{\mathbf{\dot{c}}}\,
\overline{\mathbf{R}}_{\mathbf{\dot{a}\dot{c}}}^{\mathbf{\dot{d}\dot{b}}}(u)
\,[\mathbf{\overline{x}}]_{\mathbf{\dot{d}}}^{\mathbf{b}} =
[\mathbf{\overline{ x}}]_{\mathbf{\dot{a}}}^{\mathbf{{c}}}\,
\mathbf{R}_{\mathbf{{c}a}}^{\mathbf{{b}{d}}}(u)
\,[\mathbf{x}]_{\mathbf{d}}^{\mathbf{\dot b}}
\,,\,\,\,\,\,
\overline{\mathbf{R}}_{\mathbf{\dot{a}\dot{c}}}^{\mathbf{\dot{d}\dot{b}}}(u)
\,[\mathbf{\overline{x}}]_{\mathbf{\dot{d}}}^{\mathbf{a}}
[\mathbf{\overline{x}}]_{\mathbf{\dot{b}}}^{\mathbf{c}}
=
[\mathbf{\overline{x}}]_{\mathbf{\dot{a}}}^{\mathbf{d}}
[\mathbf{\overline{x}}]_{\mathbf{\dot{c}}}^{\mathbf{b}}
\mathbf{R}_{\mathbf{{d}b}}^{\mathbf{{a}{c}}}(u)
\end{align}
which are often tacitly used in computations to move the matrix $\mathbf{R}(u)$ along paired spinorial structures. In the diagrammatic notation of Fig.\ref{fig_propag_R} they amount to the equalities:
\begin{figure}[H]
\begin{center}
\includegraphics[scale=1.8]{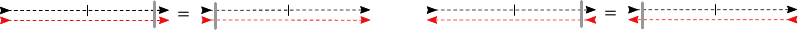}
\end{center}
\end{figure}
\noindent
With the same notation, we can represent the properties of the $\mathbf{R}$-matrix, as in Fig.\ref{unit_cross_R}.
\noindent
In the following we will write the spin numbers $n,m$ in the notation $\mathbf{R}_{nm}(u)$ only when using the compact form \eqref{compact_xp}, while we will drop them when the spinor indices are written explicitly. Let us define the matrix function
\begin{equation}
\label{S_def}
S_{\ell\dot{\ell}}(u,x) = x^{2u}\,[\mathbf{\overline{x}}]^{\ell}\mathbf{R}_{\ell\dot{\ell}}(u)
[\mathbf{x}]^{\dot{\ell}}\,,
\end{equation}
which is related to $G_{\ell\dot{\ell}}(-u-2,p)$ \eqref{prop_CFT_1} according to \eqref{Fspinor1}.

Similarly, using~\eqref{Fspinor2} it is possible to realize the operator $S_{\ell\dot{\ell}}(u,\hat p)$ as an integral operator with the kernel $G_{\ell\dot{\ell}}(-u-2,x-y)$
\begin{align}
&\hat p^{2u}\,[\mathbf{\overline{ p}}]^{\ell}\mathbf{R}_{\ell\dot{\ell}}(u)
[\mathbf{ p}]^{\dot{\ell}}\,
\Phi(x) = \frac{1}{a_{\ell\dot{\ell}}(u)}\,
\int d^4y\,
\frac{[\mathbf{\overline{x-y}}]^{\ell}\,
[\mathbf{x-y}]^{\dot{\ell}}}{(x-y)^{2(u+2)}}
\,\Phi(y)\,.
\end{align}

\begin{figure}[H]
\begin{center}
\includegraphics[scale=1.5]{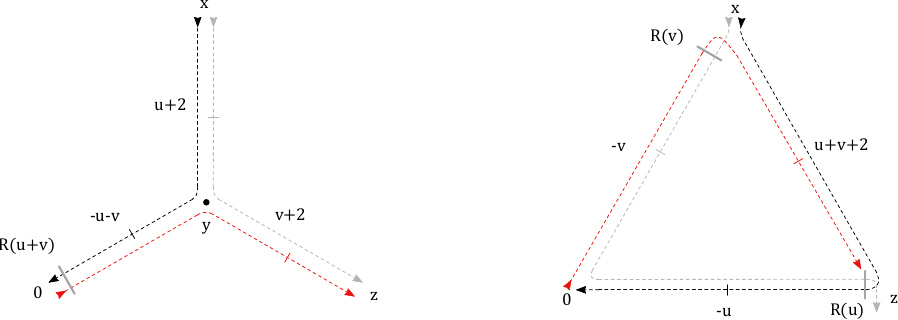}
\end{center}
\caption{\textbf{Left:} Diagrammatic form of the l.h.s. \eqref{STR_ker1} of the star-triangle duality as a \emph{star} of propagators. \textbf{Right:} Diagrammatic form of the r.h.s. \eqref{STR_ker2} of the star-triangle duality as a \emph{triangle} of propagators. The spinor indices are mixed by the $SU(2)$ fused $\mathbf{R}$-matrices that are contracted with the matrices $\boldsymbol{\sigma}, \boldsymbol{\overline{\sigma}}$ appearing in the definition \eqref{prop_CFT_1}, and their position is denoted by grey solid lines. The black dot in the star is the integrated point $y$. The letter adjacent to each segment is the scaling dimension of the propagator.}
\label{braid_STR}
\end{figure}

\begin{figure}[H]
\begin{center}
\includegraphics[scale=1.5]{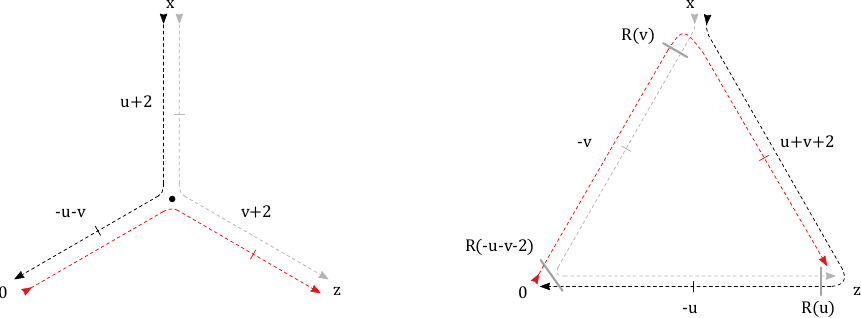}
\end{center}
\caption{Diagrammatic representation of the \emph{star-triangle}  duality \eqref{STR_exp_lhs}=\eqref{STR_exp_rhs}. Here the star of conformal propagators (l.h.s.) does not contain any mixing of spinors, which happens only in the triangle (r.h.s.). The arrows denote the order in the matrix products of $\sig,\bsig$. Grey solid lines denote $\mathbf{R}$ matrices, and the black dot in the star is the integration point. The scaling dimension of each propagator is written next to it.}
\label{STR_good}
\end{figure}

The generalization of the algebraic relation \eqref{STR_scalar_op} to spinning representations is achieved by replacing the operators $\hat p^{2u}$ and $x^{2u}$ with their spinning counterparts $S(u,p)$ and $S(u,x)$ as follows
\begin{align}
\begin{aligned}
\label{STR_operatorial}
&S_{\dot{\ell} m}(u,p) S_{m \ell}(u+v,x) S_{\ell\dot{\ell}}(v,p)=
S_{\dot{\ell}\ell}(v,x) S_{\ell m}(u+v,p) S_{m \dot{\ell}}(u,x)\,,
\end{aligned}
\end{align}
or more explicitly:
\begin{align}
\begin{aligned}
\label{STR_operatorial0}
&\hat p^{2u}\,[\mathbf{ \overline{p}}]^{\dot{\ell}}\mathbf{R}_{m\dot{\ell}}(u)
[\mathbf{p}]^{m}\,\,
x^{2(u+v)}\,[\mathbf{ \overline{x}}]^{m}\mathbf{R}_{m \ell}(u+v)
[\mathbf{x}]^{\ell}\,\,
\hat p^{2v}\,[\mathbf{\overline{ p}}]^{\ell}\mathbf{R}_{\ell\dot{\ell}}(v)
[\mathbf{p}]^{\dot{\ell}}\, = \\
&x^{2v}\,[\mathbf{ \overline{x}}]^{\dot{\ell}}\mathbf{R}_{\ell\dot{\ell}}(v)
[\mathbf{x}]^{\ell}\,\,
\hat p^{2(u+v)}\,[\mathbf{ \overline{p}}]^{\ell}\mathbf{R}_{m\ell}(u+v)
[\mathbf{p}]^{m}\,\,
x^{2u}\,[\mathbf{ \overline{x}}]^{m}\mathbf{R}_{m \dot{\ell}}(u)
[\mathbf{x}]^{\dot{\ell}}\,\,.
\end{aligned}
\end{align}
The propagators involved in the spinning star-triangle duality \eqref{STR_operatorial0} carry spin indices in the spaces
\begin{equation}
\text{Sym}_{m}[\mathbb{C}^2]\otimes \text{Sym}_{\ell}[\mathbb{C}^2]\otimes \text{Sym}_{\dot \ell}[\mathbb{C}^2]\,,
\end{equation}
which are pair-wise mixed on both the star and the triangle sides of the identity by the $\mathbf{R}$-matrix. Applying the both sides of \eqref{STR_operatorial0} to the delta-function $\delta^{(4)}(x-z)$ we obtain the following identity for integral kernels
\begin{multline}\label{STR_ker}
\int d^4 y\,\frac{[\mathbf{\overline{x-y}}]^{\dot{\ell}}
[\mathbf{x-y}]^{m}}{(x-y)^{2(u+2)}}
\frac{[\mathbf{ \overline{y}}]^{m}\mathbf{R}_{m \ell}(u+v)
[\mathbf{y}]^{\ell}}{y^{-2(u+v)}}
\frac{[\mathbf{\overline{y-z}}]^{\ell}
[\mathbf{y-z}]^{\dot{\ell}}}{(y-z)^{2(v+2)}} = \\ = \frac{a_{\dot{\ell}m}(u)a_{\ell\dot{\ell}}(v)}{a_{\ell m}(u+v)}
\frac{[\mathbf{ \overline{x}}]^{\dot{\ell}}\mathbf{R}_{\ell \dot{\ell}}(v)
[\mathbf{x}]^{\ell}}{x^{-2v}}
\frac{[\mathbf{\overline{x-z}}]^{\ell}
[\mathbf{x-z}]^{m}}{(x-z)^{2(u+v+2)}}
\frac{[\mathbf{ \overline{z}}]^{m}\mathbf{R}_{m \dot{\ell}}(u)
[\mathbf{z}]^{\dot{\ell}}}{z^{-2u}}\,.
\end{multline}
The l.h.s. of \eqref{STR_ker} is the integrated vertex $y$ of three propagators. Each of them in one the irreps $(2-v,\ell,\dot\ell)$, $(u+v,m,\ell)$ and $(2-u,\dot\ell,m)$, and it is represented on the left of Fig.\ref{braid_STR}. In explicit form it reads
\begin{align}
\begin{aligned}
\label{STR_ker1}
&\mathbf{R}(u+v)_{\mathbf{sa}}^{\mathbf{ds'}} \int d^4y\, \frac{[(\mathbf{x-y})\mathbf{\overline{y}}]_{\mathbf{b}}^{\mathbf{s}} [\mathbf{y(\overline{y-z})}]_{\mathbf{s'}}^{\mathbf{c}}
[\mathbf{(\overline{x-y})(y-z)}]_{\mathbf{\dot a}}^{\mathbf{\dot{d}}}}
{(x-y)^{2(u+2)}y^{2(-u-v)}(y-z)^{2(v+2)}}\,.
\end{aligned}
\end{align}
Repeated spinorial indices are contracted, and the indices in $\text{Sym}_{\ell}\otimes \text{Sym}_m$ are mixed by the matrix $\mathbf{R}_{\ell m}(u+v)$.  The \emph{star} vertex is scale invariant as the sum of scaling dimensions of the propagators is equal to the dimension $d=4$ of the space and is proportional to a product of propagators in the irreps $(u,\ell,\dot\ell)$, $(2-u-v,m,\ell)$ and $(v,\dot\ell,m)$, forming the \emph{triangle} in the right of Fig.\ref{braid_STR}.  On the r.h.s. of Fig.\ref{braid_STR} the indices in $\text{Sym}_{\dot\ell}\otimes \text{Sym}_m$ and $\text{Sym}_{\ell}\otimes \text{Sym}_{\dot\ell}$ are mixed pairwise by the matrices $\mathbf{R}_{m\dot{\ell}}(u)$ and $\mathbf{R}_{\ell\dot{\ell}}(v)$
\begin{align}
\begin{aligned}
\label{STR_ker2}
&\mathbf{R}(u)_{\mathbf{ec'}}^{\mathbf{de'}}\mathbf{R}(v)_{\mathbf{ra}}^{\mathbf{c'r'}}\,
\frac{[\mathbf{\overline{x}}]_{\mathbf{\dot{a}}}^{\mathbf{r}}
[\mathbf{x}\mathbf{\overline{(x-z)}}]_{\mathbf{r'}}^{\mathbf{c}}
[(\mathbf{x-z})\mathbf{\overline{z}}]_{\mathbf{b}}^{\mathbf{e}}
[\mathbf{z}]_{\mathbf{e'}}^{\mathbf{\dot{d}}}}{{{x}^{2(-v)}(x-z)^{2(2+u+v)}{z}^{2(-u)}}} \,,
\end{aligned}
\end{align}
\noindent
We can cast the identity in a more convenient form - depicted in Fig.\ref{STR_good} - where the mixing of spinors is removed from the l.h.s. of \eqref{STR_ker},  moving the matrix $\mathbf{R}(u+v)$ to the r.h.s by means of crossing symmetry \eqref{crossinxg}. Indeed, contracting \eqref{STR_ker1} and \eqref{STR_ker2} with the matrix $\mathbf{R}(-u-v-2)_{\mathbf{d a'}}^{\mathbf{d'a}}$ one cancels any mixing on the l.h.s. of the equation. After relabeling $\mathbf{a'}\to \mathbf{a}$, $\mathbf{d'}\to \mathbf{d}$ the star integral takes the form
\begin{align}
\begin{aligned}
\label{STR_exp_lhs}
& \int d^4y\,
\frac{[(\mathbf{x-y})\mathbf{\overline{y}}]_{\mathbf{b}}^{\mathbf{d}} [\mathbf{y(\overline{y-z})}]_{\mathbf{a}}^{\mathbf{c}}
[\mathbf{(\overline{x-y})(y-z)}]_{\mathbf{\dot a}}^{\mathbf{\dot{d}}}}
{(x-y)^{2(u+2)}y^{2(-u-v)}(y-z)^{2(v+2)}}\,,
\end{aligned}
\end{align}
while the triangle on the r.h.s. of the equation is
\begin{align}
\begin{aligned}
\label{STR_exp_rhs}
&\frac{a_{\dot{\ell}m}(u)a_{\ell\dot{\ell}}(v)}
{r_{m\ell}(u+v)a_{\ell m}(u+v)}\mathbf{R}(-u-v-2)_{\mathbf{sa}}^{\mathbf{ds'}}
\mathbf{R}(u)_{\mathbf{ec'}}^{\mathbf{se'}}
\mathbf{R}(v)_{\mathbf{rs'}}^{\mathbf{c'r'}}\,
\frac{[\mathbf{\overline{x}}]_{\mathbf{\dot{a}}}^{\mathbf{r}}
[\mathbf{x}\mathbf{\overline{(x-z)}}]_{\mathbf{r'}}^{\mathbf{c}}
[(\mathbf{x-z})\mathbf{\overline{z}}]_{\mathbf{b}}^{\mathbf{e}}
[\mathbf{z}]_{\mathbf{e'}}^{\mathbf{\dot{d}}}}{{{x}^{2(-v)}(x-z)^{2(2+u+v)}{z}^{2(-u)}}}
\,.
\end{aligned}
\end{align}
In this form, illustrated in Fig.\ref{STR_good}, the star-triangle identity relates a conformal-invariant vertex of three spinning propagators of type $G(u,x)$ with the product of three propagators whose spinorial indices are mutually mixed by the $\mathbf{R}$-matrix.
%\begin{multline}
%\label{STR_exp_lhs0}
%\int d^4y\,
%\frac{[(\mathbf{x-y})\mathbf{\overline{y}}]_{\mathbf{b}}^{\mathbf{d}} [\mathbf{y(\overline{y-z})}]_{\mathbf{a}}^{\mathbf{c}}
%[\mathbf{(\overline{x-y})(y-z)}]_{\mathbf{\dot a}}^{\mathbf{\dot{d}}}}
%{(x-y)^{2(u+2)}y^{2(-u-v)}(y-z)^{2(v+2)}}\, = \frac{n_{\ell m}(u+v)}
%{r_{m\ell}(u+v)n_{\dot{\ell}m}(u)n_{\ell\dot{\ell}}(v)}\,\times \\
%\times\,
%\mathbf{R}(-u-v-2)_{\mathbf{sa}}^{\mathbf{ds'}}
%\mathbf{R}(u)_{\mathbf{ec'}}^{\mathbf{se'}}
%\mathbf{R}(v)_{\mathbf{rs'}}^{\mathbf{c'r'}}\,
%\frac{[\mathbf{\overline{x}}]_{\mathbf{\dot{a}}}^{\mathbf{r}}
%[\mathbf{x}\mathbf{\overline{(x-z)}}]_{\mathbf{r'}}^{\mathbf{c}}
%[(\mathbf{x-z})\mathbf{\overline{z}}]_{\mathbf{b}}^{\mathbf{e}}
%[\mathbf{z}]_{\mathbf{e'}}^{\mathbf{\dot{d}}}}{{{x}^{2(-v)}(x-z)^{2(2+u+v)}{z}^{2(-u)}}}
%\end{multline}
%For clarity we present its compact form
%\begin{multline}\label{STR_ker0}
%\int d^4 y\,\frac{
%[\mathbf{(x-y)}\mathbf{\overline{y}}]^{m}
%[\mathbf{y}\mathbf{\overline{(y-z)}}]^{\ell}
%[\mathbf{\overline{(x-y)}}\mathbf{(y-z)}]^{\dot{\ell}}
%}{(x-y)^{2(u+2)}y^{-2(u+v)}(y-z)^{2(v+2)}}
% = \frac{n_{\ell m}(u+v)}
%{r_{m\ell}(u+v)n_{\dot{\ell}m}(u)n_{\ell\dot{\ell}}(v)}\,\times
%\\
%\times\frac{\tr_m \mathbf{R}_{m\ell}(-u-v-2)
%[\mathbf{ \overline{x}}]^{\dot{\ell}}
%\mathbf{R}_{\ell \dot{\ell}}(v)
%[\mathbf{x}\mathbf{\overline{(x-z)}}]^{\ell}
%[\mathbf{(x-z)}\mathbf{ \overline{z}}]^{m}
%\mathbf{R}_{m \dot{\ell}}(u)
%[\mathbf{z}]^{\dot{\ell}}}{x^{-2v}(x-z)^{2(u+v+2)}z^{-2u}}\,.
%\end{multline}
%where $\tr_m$ is exactly means summation over index $\mathbf{s}$ in the previous formula.
\noindent
Let us point out that the diagrammatic approach of figures \ref{scalar_STR}-\ref{STR_good} is strikingly simpler than the analytic one, which motivates our artistic efforts in this section. The involved structure of spinorial indices and their contractions, for instance in \eqref{STR_ker}-\eqref{STR_exp_rhs}, is entirely contained in the notation of two oriented dashed lines and solid gray $\mathbf{R}$-matrix insertions. In fact, the diagrammatic approach is the most compact way to perform computations in conformal quantum mechanics without loss of rigour, hence in most of the following we will avoid any analytic formula in favour of graphical derivations. %Any computation based on \eqref{STR_ker} is definitely simpler to do at the diagrammatic level as a sequence of graphical transformations, just keeping track of the labels $a,b,c$ in order to reconstruct the coefficient produced at each application of the duality.   Among the several equivalent forms of the identity \eqref{STR_operatorial},  we present in detail the most useful for direct application to  the spin chain computations of the next sections. First, we must mention the chain-rule identity resulting from the amputation of one propagator in the l.h.s. of \eqref{STR_most_symm}, that is by sending the point $x_1$ to $\infty$ via a conformal inversion
%\enrico{check position R-matrices}
%\begin{align}
%\label{Chain_rule}
%\begin{aligned}
%&\int d^4 y \frac{[\mathbf{(y-x_3)}]^{{\ell}}[\mathbf{(\overline{y-x_2})}]^{\dot{n}}[\mathbf{(x_2-y)(\overline{y-x_3})}]^{\dot{\ell}}}{(x_1-y)^{2(a)}(x_2-y)^{2(b)}(x_3-y)^{2(c)}}=\\&=\int D\tau \frac{\langle \tau |\mathbf{R}_{\ell\dot{\ell}}(c-2)[\mathbf{x_{23}}]^{{\ell}}\mathbf{R}_{\ell\dot{n}}(a-2)[\mathbf{\overline{x_{32}}}]^{\dot{n}}\mathbf{R}_{\dot{\ell}\dot{n}}(b-2)|\tau\rangle}{(x_{23})^{2(2-a)}(x_{13})^{2(2-b)}(x_{12})^{2(2-c)}}\\&\times\frac{\pi^2(-1)^{\ell}}{\left(a-1+\frac{\ell+\dot{n}}{2}\right)\left(c-1+\frac{\dot{\ell}+\ell}{2}\right)\left(b-1+\frac{\dot{n}+\dot{\ell}}{2}\right)} \frac{\Gamma\left(2-a+\frac{\dot{n}-\ell}{2}\right)\Gamma\left(2-c+\frac{\ell-\dot{\ell}}{2}\right)\Gamma\left(2-b+\frac{\dot{n}-\dot{\ell}}{2}\right)}{\Gamma\left(c-1+\frac{\ell-\dot{\ell}}{2}\right)\Gamma\left(b-1+\frac{\dot{n}-\dot{\ell}}{2}\right)\Gamma\left(a-1+\frac{\dot{n}-\ell}{2}\right)}\,.
%\end{aligned}
%\end{align}
%\begin{figure}[H]
%\begin{center}
%\includegraphics[scale=1.5]{Chain_rule}
%\end{center}
%\caption{Diagrammatic representation of the chain-rule identity \eqref{Chain_rule}. One of spinor structures in the l.h.s. (colored in red) disappears as a result of the integration, and its indices are left on the two R-matrices $\mathbf{R}(2-u-v)$ and $\mathbf{R}(v)$ }
%\label{chain_rule_diag}
%\end{figure}
%\noindent

Among the integral identities that one can derive by means of \eqref{STR_operatorial} we present two which are highly frequent in the computations of the next sections, which we dub \emph{interchange relations}. Both identities describe the transformation of a conformal-invariant vertex of four spinning propagators (see Fig.\ref{Spinning_fish}) by a reshuffling of scaling dimensions and spins of the propagators. The first identity holds under the constraint $u+u'=v+v'$ and has the diagrammatic form:
\begin{figure}[H]
\begin{center}
\includegraphics[scale=1.3]{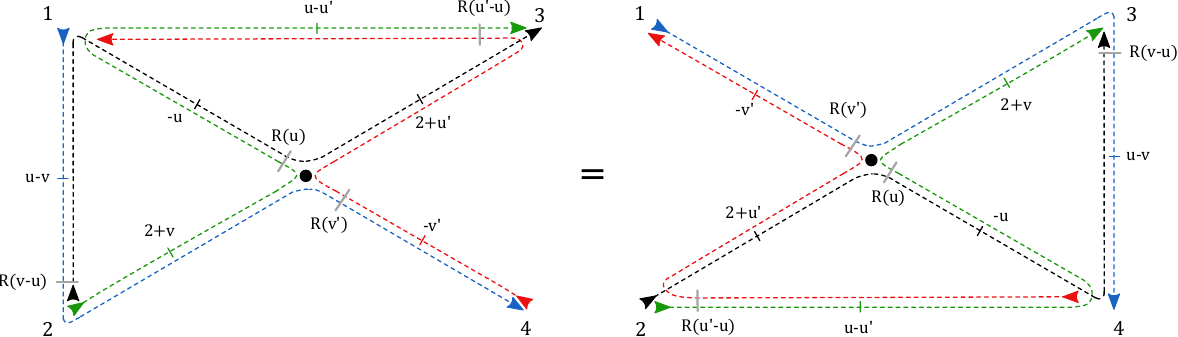}
\end{center}
\caption{Diagrammatic form of the interchange relation I.}
\label{Inter_I}
\end{figure}
\noindent
where the equality between the two diagrams is exact, i.e. the proportionality factor is $1$. The proof of the identity consists of four star-triangle steps
\begin{itemize}
\item The black/blue line $12$ in the l.h.s of Fig.\ref{Inter_I} is the edge of a triangle of propagators whose scaling dimensions sum to $2$, thus it can be rewritten as a star integral of vertices $\{1,2,\bullet\}$ as in Fig.\ref{braid_STR}.
\item The point of $\bullet$ is now the integrated point of a star integral of propagators whose scaling dimensions sum to $4$. Therefore it can be integrated delivering a triangle whose basis is a black/blue line $34$, again as in Fig.\ref{braid_STR}. The move of the line $12$ to $34$ is completed.
\item The same two steps can be performed on the red/green line $13$, which in turn gets moved downstairs to $24$, resulting in the r.h.s. of Fig.\ref{Inter_I}.
\end{itemize}
The second identity differs from the first only in the mutual orientation of spinorial lines and position of $\mathbf{R}(u)$-matrices. The proof follows the very same steps as for the first relation, adapted only in the variant of star-triangle identity involved, and its diagrammatic form is:
\begin{figure}[H]
\begin{center}
\includegraphics[scale=1.3]{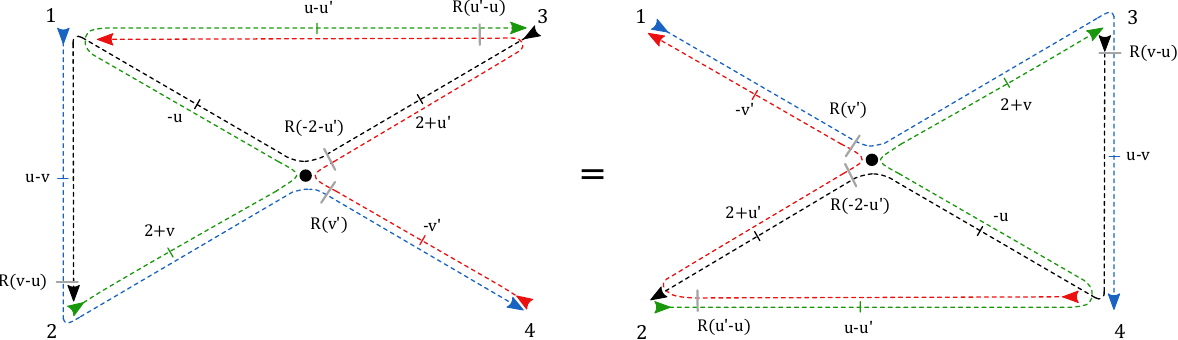}
\end{center}
\caption{Diagrammatic form of the interchange relation II.}
\label{Inter_II}
\end{figure}
\noindent
The statement of the interchange relations can be read out of Fig.\ref{Inter_I} and Fig.\ref{Inter_II}: moving the horizontal (red/green) propagator and the vertical (grey/black) propagator across the quartic scale invariant vertex, the powers $(u,v,u',v')$ in the vertex propagators get interchanged, and so do the spinorial structures depicted by different colors of lines.

\section{General $SO(1,5)$ R-operator}
\label{sect:gen}
In this section we present a solution of the Yang-Baxter equation for the group $SO(1,5)$ in the unitary irreps of the principal series. As the representation is infinite-dimensional, the solution is an infinite dimensional $R$-operator. Its derivation is based on the star-triangle duality, and we realize it in the diagrammatic notation of Feynman integrals. The knowledge of an $R$-operator is the main object needed to formulate the quantum integrability of a system of interacting particles according to the QISM approach \cite{Faddeev:1996iy,Kulish:1981bi}, and it will be the topic of the next sections.

An $R$-operator $\rmat_{ij}(u)$ is an endomorphism over the tensor product of two representation modules $\mathbb{V}_i \otimes \mathbb{V}_j$ that depends on the \emph{spectral parameter} $u\in \mathbb{C}$,  and solves the Yang-Baxter equation (YBE)~\cite{Kulish:1980ii}
\begin{equation}
\label{YBE_inf}
\rmat_{12}(u-v)\rmat_{13}(u)\rmat_{23}(v) \,=\,\rmat_{23}(v) \rmat_{13}(u) \rmat_{12}(u-v)\,,
\end{equation}where the action of $\rmat_{ij}(u)$ on any module $\mathbb{V}_{k}$,
where $k\neq i,j$ is trivial.  The module $\mathbb{V}_k$ is the Hilbert space of the $k$-th particle of the system, i.e. the space of functions of the point $x^{\mu}\in \mathbb{R}^4$ with symmetric spinor indices of dotted/undotted type
\begin{equation}
\label{Phi_funct}
\Phi_{\mathbf{a\dot{a}}}(x)\equiv \Phi_{(a_1,\dots,a_{\ell})\,(\dot{a}_1,\dots, \dot{a}_{{\ell}})}(x)\,,
\end{equation}
and it carries an infinite-dimensional irrep of $SO(1,5)$ labelled by the scaling dimension $\Delta\,\in\,\mathbb{C}$ and the left/right spins $(\ell,\dot{\ell})$
\begin{equation}
\mathbb{V}_k\, \longleftrightarrow\, (\Delta_k,\ell_k,\dot{\ell}_k)\,.
\end{equation}
Explicitly, the action of the $R$-operator reads
\begin{equation}
\Phi(x_1)_{\mathbf{a\dot{a}}}\Phi(x_2)_{\mathbf{b\dot{b}}}\,\mapsto\, \Phi'(u;x_1,x_2)_{\mathbf{a\dot{a}b\dot{b}}} =\left(\rmat_{12}(u)\right)_{\mathbf{a\dot{a}b\dot{b}}}^{\mathbf{c\dot{c}d\dot{d}}} \Phi(x_1)_{\mathbf{c\dot{c}}}\Phi(x_2)_{\mathbf{d\dot{d}}}\,.
\end{equation}
The function $\Phi'(u,x_1,x_2)$ must belong to the tensor product of representations $\mathbb{V}_1\otimes\mathbb{V}_2$,  implying that a solution of YBE must be conformal invariant
\begin{equation}
\label{confinv}
( T_1 \otimes T_2) \rmat_{12}(u)\ = \rmat_{12}(u)( T_1 \otimes T_2)\,,
\end{equation}
for $T_k$ representative of a $SO(1,5)$ group element in the representation $(\Delta_k,\ell_k,\dot{\ell}_k)$.

A solution $\rmat_{ij}(u)$ of the YBE can be built starting from two observations~\cite{Chicherin:2012yn}.
First,  it is always possible to formulate the problem in an equivalent
way as a braid equation
\begin{equation}
\label{braid_R}
\hat \rmat_{12}(u)\hat \rmat_{23}(u+v)\hat \rmat_{12}(v)=\hat \rmat_{23}(v)\hat \rmat_{12}(u+v)\hat \rmat_{23}(u)\,,
\end{equation}
where $\hat{\rmat}_{ij}(u)=\mathbb{P}_{ij}\rmat_{ij}(u)$ and $\mathbb{P}_{ij} (\mathbb{V}_i \otimes \mathbb{V}_j)=  \mathbb{V}_j \otimes \mathbb{V}_i$.  Secondly, the star-triangle duality \eqref{STR_operatorial} can be considered as a braid relation itself,
involving operators that act on the spaces of two particles $\mathbb{V}_1\otimes \mathbb{V}_2$
\begin{align}\label{int1}
&S_{\ell_k\dot{\ell}_k}(u,\hat p_k) =
\hat p_k^{2u}\,[\mathbf{\overline{p}}_k]^{\ell_k}\mathbf{R}_{\ell_k\dot{\ell}_k}(u)
[\mathbf{p}_k]^{\dot{\ell}_k}\,,\,\,\,\,k=1,2\,,\\\label{int2}&S_{\ell_1\dot{\ell}_2}(u,x_{12}) = x_{12}^{2u}\,[\mathbf{\overline{x}}_{12}]^{\ell_1}
\mathbf{R}_{\ell_1\dot{\ell}_2}(u)
[\mathbf{x}_{12}]^{\dot{\ell}_2}\,.
\end{align}
%taking the form
%\begin{equation}
%label{STR_12_op}
%S(u,\hat p_k)S(u+v,x_{12})S(v,\hat p_k)=S_{v}(x_{12})S_{u+v}(\hat p_k)S(u,x_{12})\,,\,\,\,\,k=1,2\,.
%end{equation}
Hence, it is natural to look for a solution of \eqref{braid_R} as a combination $\hat\rmat_{12}(u)$ of the operators $S(u,x_{12})$ and $S(u,p_k)$ that inherits the braid property from the star-triangle duality and exchanges the representations $(\Delta_1,\ell_1,\dot{\ell}_1)$ and $(\Delta_2,\ell_2,\dot{\ell}_2)$ of the modules $\mathbb{V}_1\otimes \mathbb{V}_2$
\begin{align}
\label{R_hat_maps}
\begin{aligned}
( T_2 \otimes T_1) \hat \rmat_{12} =\hat \rmat_{12} ( T_1 \otimes  T_2)\,,
\end{aligned}
\end{align}
 as it follows from \eqref{confinv} and from the definition of $\hat{\rmat}_{12}$.  Such a combination exists and reads
\begin{align}
\label{braid_sol}
\hat\rmat_{12}(u) =
S_{\dot{\ell}_1\ell_2}\left(u-\Delta_{+},x_{12}\right)\,
S_{\ell_2\ell_1}\left(u+\Delta_{-},\hat p_2\right)\,
S_{\dot{\ell}_2\dot{\ell}_1}\left(u-\Delta_{-},\hat p_1\right)\,
S_{\ell_1\dot\ell_2}\left(u+\Delta_{+},x_{12}\right)\,,
\end{align}
or in explicit form
\begin{multline}
\label{braid_solexp}
\hat\rmat_{12}(u) = x_{12}^{2\left(u-\Delta_{+}\right)}\,
[\mathbf{\overline{x}}_{12}]^{\dot{\ell}_1}
\mathbf{R}_{\dot{\ell}_1\ell_2}(u-\Delta_{+})
[\mathbf{x}_{12}]^{\ell_2}\,\,
\hat{p}_{2}^{2(u+\Delta_{-})}\,[\mathbf{\overline{p}}_{2}]^{\ell_2}
\mathbf{R}_{\ell_2\ell_1}(u+\Delta_{-})
[\mathbf{p}_{2}]^{\ell_1}\,\,\times \\
\times\,\,
\hat{p}_{1}^{2(u-\Delta_{-})}\,[\mathbf{\overline{p}}_{1}]^{\dot{\ell}_2}
\mathbf{R}_{\dot{\ell}_2\dot{\ell}_1}(u-\Delta_{-})
[\mathbf{p}_{1}]^{\dot{\ell}_1}\,\,
x_{12}^{2\left(u+\Delta_{+}\right)}\,
[\mathbf{\overline{x}}_{12}]^{\ell_1}\mathbf{R}_{\ell_1\dot{\ell}_2}(u+\Delta_{+})
[\mathbf{x}_{12}]^{\dot{\ell}_2}
\,,
\end{multline}
where $\Delta_{-}=\frac{\Delta_1-\Delta_2}{2}$ and
$\Delta_{+}=\frac{\Delta_1+\Delta_2}{2}-2$. The combination \eqref{braid_sol} is the straightforward generalization, in full analogy to \eqref{STR_operatorial} and \eqref{STR_scalar_op},
of the spinless solution ~\cite{Chicherin:2012yn}
\begin{equation}\label{spinless}
\hat \rmat_{12}(u) =
x_{12}^{2(u-\Delta_{+})}
\hat p_1^{2(u-\Delta_{-})}
\hat p_2^{2(u+\Delta_{-})}
x_{12}^{2(u+\Delta_{+})} \,.
\end{equation}
Let us analyze the behavior of the operators \eqref{int1} and \eqref{int2} under conformal transformations.  These operators are not conformal invariants, nevertheless they define an intertwining of principal series representations, namely they map a function on $\mathbb{V}_1\otimes \mathbb{V}_2$ to the product of modules $\mathbb{V}_1'\otimes \mathbb{V}_2'$, carrying a different representations of the principal series \cite{Gurdogan:2015csr}. In particular, for the value $u=2-\Delta$ the operator $S_{\ell\dot{\ell}}(2-\Delta,\hat p)$ defines the map (shadow transformation) $(\Delta,\ell,\dot{\ell})\to\left(4-\Delta,\dot{\ell},\ell\right)$, that is intertwining relation
\begin{equation}
\label{inter_p}
T'\, S_{\ell\dot{\ell}}(2-\Delta, \hat p) = S_{\ell\dot{\ell}}(2-\Delta, \hat p)\, T\,,
\end{equation}
Note that the shadow transformation is an involution, that is expressed by the compatibility relation
\begin{align}\label{reflect1}
S_{\dot{\ell}\ell}(-u, \hat p)\,S_{\ell\dot{\ell}}(u, \hat p) = \mathbbm{1}\,,
\end{align}
and holds as a consequence of the $\mathbf R$-matrix parity \eqref{unitar}.

In the case of $S_{\ell_1\dot\ell_2}(u,x_{12})$ the transformation
involves both modules at the same time
\begin{align}
\label{inter_x}
\begin{aligned}
&(\Delta_1,{\ell}_1,\dot{\ell}_1)\otimes(\Delta_2,{\ell}_2,\dot{\ell}_2) \rightarrow \left(\Delta_1-u,\dot{\ell}_2,\dot{\ell}_1\right)\otimes
\left(\Delta_2-u,{\ell}_2,{\ell}_1\right)\,,\\& ( T'_1 \otimes T'_2 ) S_{\ell_1\dot{\ell}_2}(u,x_{12})= S_{\ell_1\dot{\ell}_2}(u,x_{12})
( T_1\otimes  T_2)\,,
\end{aligned}
\end{align}
and similarly to ~\eqref{reflect1} the involutivity is expressed by the compatibility relation
\begin{align}\label{reflect2}
S_{\dot{\ell}_2\ell_1}(-u, x_{12})\,S_{\ell_1\dot{\ell}_2}(u, x_{12}) = \mathbbm{1}\,.
\end{align}
Fixing $v=2-\Delta_1$ in \eqref{STR_operatorial}, the two sides of the star-triangle duality
\begin{multline}
\label{Braid_map}
S_{\dot{\ell}_2\ell_1}(u,\hat p_1)\,S_{\dot{\ell}_1\dot{\ell}_2}(2+u-\Delta_1,x_{12})\,
S_{\ell_1\dot{\ell}_1}(2-\Delta_1,\hat p_1) = \\
= S_{\dot{\ell}_1\ell_1}(2-\Delta_1,x_{12})\,
S_{\dot{\ell}_2\dot{\ell}_1}(2+u-\Delta_1,\hat p_1)\,
S_{\ell_1\dot{\ell}_2}(u,x_{12})\,,
\end{multline}
can be regarded as two equivalent ways to compose operators
\eqref{int1} and \eqref{int2} to realize the map between representations:
\begin{align}
\label{S_braid_mapping}
\begin{aligned}
&(\Delta_1,{\ell}_1,\dot{\ell}_1)\otimes(\Delta_2,\ell_2,\dot{\ell}_2) \longrightarrow \left(u+2,{\ell}_1,\dot{\ell}_2\right)\otimes\left(\Delta_1+\Delta_2-u-2,{\ell}_2,\dot{\ell}_1\right)\,.
\end{aligned}
\end{align}
In this respect the duality \eqref{Braid_map} together with the properties
\eqref{reflect1} and \eqref{reflect2} are the Coxeter relations for the
intertwiners of irreps of $SO(1,5)$
(for a detailed discussion see~\cite{Chicherin:2012yn}).

\begin{figure}[H]
\begin{center}
\includegraphics[scale=1.5]{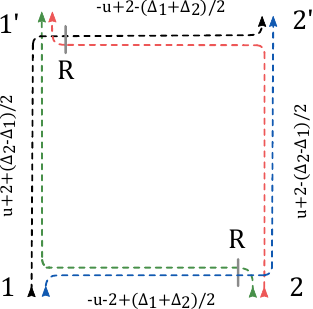}$\qquad\qquad\qquad$\includegraphics[scale=1.5]{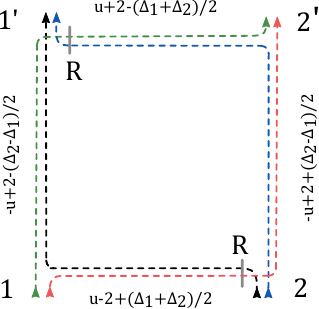}
\end{center}
\caption{\textbf{Left:}  diagram of the kernel $\hat \rmat_{12}(x_1,x_2|x_1',x_2')$ of the operator $\hat \rmat_{12}(u)$. The superscripts of each segment are the powers of the inverse square distance between the end points. Grey solid lines represent the fused $SU(2)$ R-matrix acting along the spinorial structure of the propagators. The argument of matrices $\mathbf{R}$ are the opposite of the power of the adjancent line. \textbf{Right:} diagram of the kernel of the operator $\hat \rmat^{-1}_{12}(u)$.}
\label{GEN_R}
\end{figure}

%A consequence of \eqref{braid_sol} and \eqref{reflect}\,,\eqref{reflect1} is that $
%\hat \rmat(-u)_{21}\hat \rmat(u)_{12}\,=\, \mathbbm{1}$, hence the operator $\rmat_{12}(u)$ is a triangular solution of the Yang-Baxter equation
%\begin{equation}
%\rmat_{12}(u)^{-1}=\rmat_{12}(-u)\,.
%\end{equation}
According to the definition \eqref{braid_sol},  the solution of the braid relation $\hat\rmat_{12}$ acts on functions of the space coordinates $x_1$ and $x_2$ as an integral operator
\begin{align}
\begin{aligned}
\label{R_integral_form}
\hat\rmat_{12}(u)\,\Phi(x_1\,,x_2)  =
\frac{1}{a_{\dot{\ell}_2\dot{\ell}_1}(u-\Delta_{-})\,
a_{\ell_2\ell_1}(u+\Delta_{-})}\,
\int d^4x'_1 d^4x'_2 \,\hat \rmat(x_1,x_2|x'_1,x'_2)\,\Phi(x'_1\,,x'_2)\,.
\end{aligned}
\end{align}
Its kernel $\hat\rmat(x_1,x_2|x'_1,x'_2)$ in compact notation reads
\begin{align*}
\frac{[\mathbf{\overline{x}}_{12}]^{\dot{\ell}_1}
\mathbf{R}_{\dot{\ell}_1\ell_2}(u-\Delta_{+})
[\mathbf{x}_{12}]^{\ell_2}}{x_{12}^{2\left(-u+\Delta_{+}\right)}}
\frac{[\mathbf{\overline{x}}_{22'}]^{\ell_2}
[\mathbf{x}_{22'}]^{\ell_1}}
{x_{22'}^{2\left(u+\Delta_{-}+2\right)}}
\frac{[\mathbf{\overline{x}}_{11'}]^{\dot{\ell}_2}
[\mathbf{x}_{11'}]^{\dot{\ell}_1}}
{x_{11'}^{2\left(u-\Delta_{-}+2\right)}}
\frac{[\mathbf{\overline{x}}_{1'2'}]^{\ell_1}\mathbf{R}_{\ell_1\dot{\ell}_2}(u+\Delta_{+})
[\mathbf{x}_{1'2'}]^{\dot\ell_2}}
{x_{1'2'}^{2\left(-u-\Delta_{+}\right)}}\,,
\end{align*}
and as a diagram it is a square of propagators with spinor mixing, as depicted in Fig.\ref{GEN_R}.
Using \eqref{reflect1} and \eqref{reflect2} it is easy to construct
inverse operator $\hat \rmat^{-1}_{12}(u)$ and its kernel is shown in the same Fig.\ref{GEN_R}.

\begin{figure}[H]
\begin{center}
\includegraphics[scale=1.22]{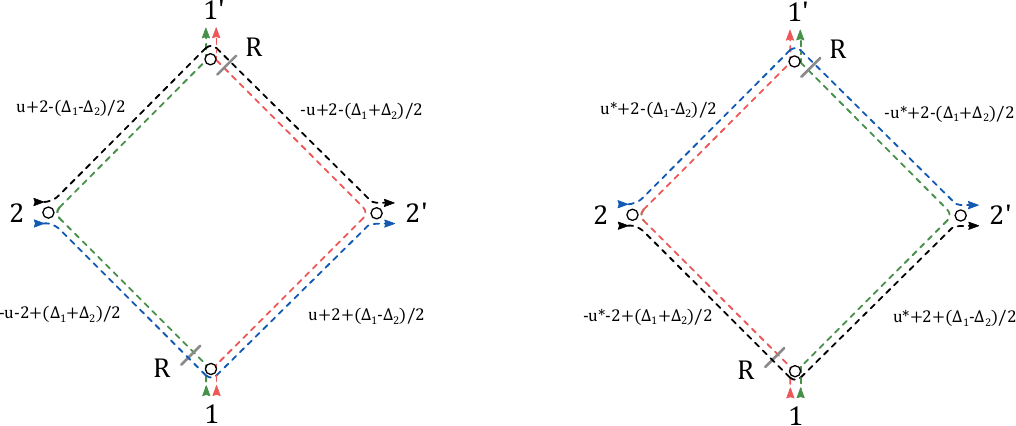}
\end{center}
\caption{\textbf{Left:} diagrammatic representation of the kernel $\rmat_{12}(x_1,x_2|x_1',x_2')$ of the operator $\rmat_{12}(u)$. It is obtained from the kernel of Fig.\ref{GEN_R} (left) by exchange of the points $1$ and $2$ following from the definition $\rmat_{12}=\mathbb{P}_{12}\hat\rmat_{12}$. \textbf{Right:} diagrammatic representation of the kernel of the operator $\rmat^{\dagger}_{12}(u)$. The color of lines - that is the values of the spins - get exchanged by the hermitean conjugation.
}
\label{R_mat_YBE}
\end{figure}

\noindent
The operator $\rmat_{12}(u)$ enjoys a simple behavior under hermitean conjugation; the form of the YBE ensures that $\rmat^{\dagger}_{12}(u)$ is still a solution.
For the $S$-operator we have
\begin{align*}
S^{\dagger}_{\ell\dot{\ell}}(u,\hat{p}) = \hat p^{2u^*}\,
[\mathbf{\overline{p}}]^{\dot{\ell}}
\mathbf{R}_{\ell\dot{\ell}}(u^*)[\mathbf{p}]^{\ell} =
S_{\dot{\ell}\ell}(u^*,\hat{p}) = S^{-1}_{\ell\dot{\ell}}(-u^*,\hat{p})\,,
\end{align*}
and the same in coordinate representation.
Then from the definition \eqref{braid_sol} it follows that
\begin{equation}
\label{R_hc}
\rmat^{\dagger}_{12}(u,\Delta_{\pm}) \,=\,\rmat^{-1}_{12}(-u^*,-\Delta^*_{\pm})\,.
\end{equation}
where we show explicitly dependence on the parameters $\Delta_{\pm}$ in $\rmat$-operator.
For the unitary series representations of conformal group parameters are fixed as follows:
$\Delta_1 = 2+i\nu_1\,,\Delta_2 = 2+i\nu_2$ so that
$\Delta_{-} = \frac{i}{2}(\nu_1-\nu_2)$ and $\Delta_{+} = \frac{i}{2}(\nu_1+\nu_2)$
are pure imaginary; in this case for imaginary spectral parameter $\rmat_{12}(u)$ is a unitary operator. The integral kernels of the operators $\rmat_{12}(u)$,
$\rmat^{\dagger}_{12}(u)$ are drawn in Fig.\ref{R_mat_YBE}.

Since the Yang-Baxter equation for $\rmat_{ij}(u)$ is ultimately a consequence of the braid relations satisfied by the operators \eqref{int1} and \eqref{int2},  its proof can delivered in the diagrammatic formalism developed in section \ref{cqm_str}, with few transparent steps. For a choice of three particles/modules in any unitary irreps \begin{align*}
&\mathbb{V}_1 \,\to\, (\Delta_1,\ell_1,\dot{\ell}_1)\,,\,\,\,\,\mathbb{V}_2 \,\to\, (\Delta_2,\ell_2,\dot{\ell}_2)\,,\,\,\,\,\mathbb{V}_3 \,\to\, (\Delta_3,\ell_3,\dot{\ell}_3)\,,
\end{align*}
the l.h.s. of \eqref{R_mat_YBE} is a convolution of the integral kernels of $\rmat_{ij}$ operators (left picture). Here each color denotes spinorial indices - carried by matrices $\boldsymbol{\sigma}$, $\boldsymbol{\bar{\sigma}}$ - belonging to different spinor spaces. The first step in the proof consist in redrawing the triangle $(y_1,y_2,y_3)$ as a star integral via the identity \eqref{STR_ker} (right picture)
\begin{center}
\includegraphics[scale=0.85]{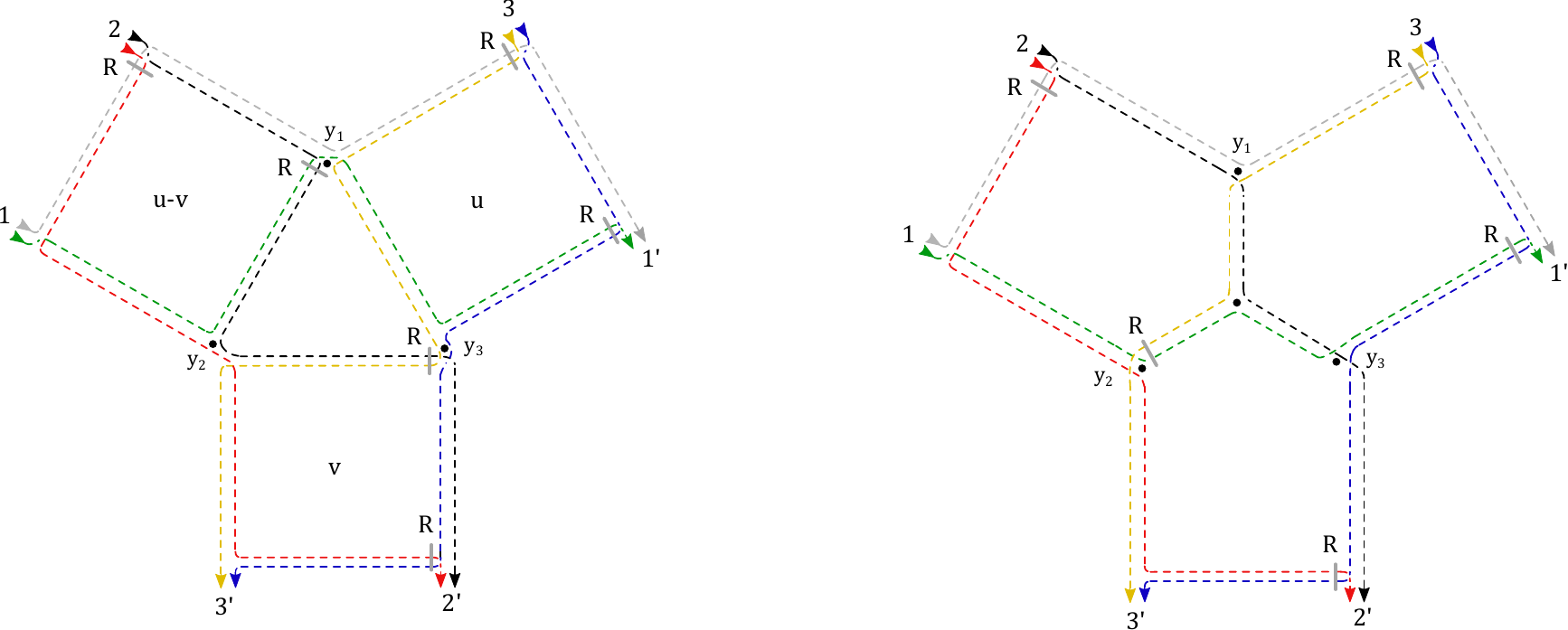}
\end{center}
At the second step, we apply the same identity \eqref{STR_good} to integrate the scale invariant vertices $y_1$, $y_2$ and $y_3$, transforming each star into a triangle, and obtaining the symmetric picture:
\begin{center}
\includegraphics[scale=0.8]{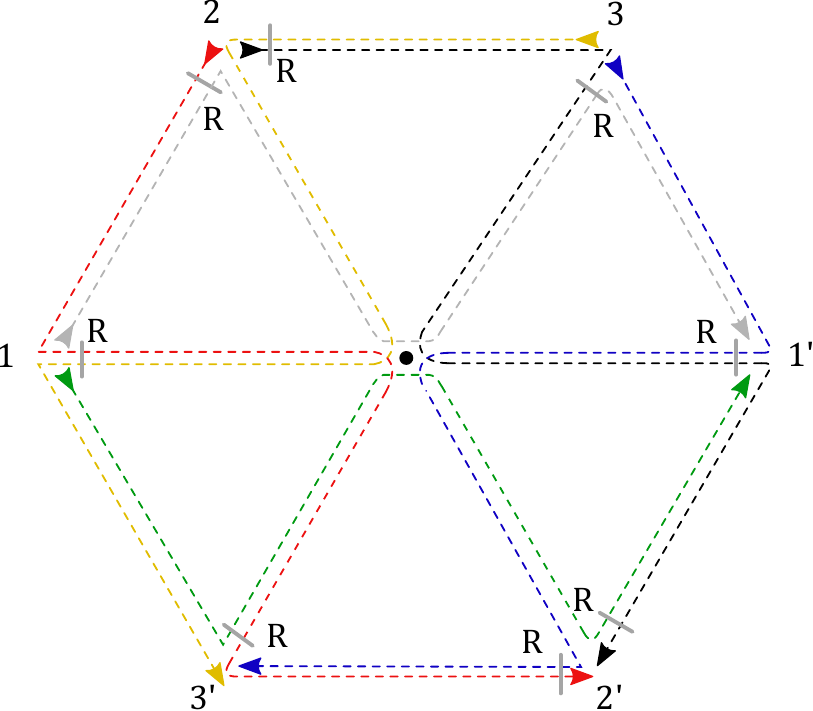}
\end{center}
The same two steps can now be repeated starting from the r.h.s. of the \eqref{R_mat_YBE}, which amounts to a different order in the convolution of kernels ${\rmat}$.
%\begin{center}
%\includegraphics[scale=0.8]{Gen_YBE_2}
%\end{center}
It is straightforward to check that as a final step one obtains the same sextic vertex of propagators, proving the Yang-Baxter property $\square$.

\begin{figure}[H]
\begin{center}
\includegraphics[scale=0.85]{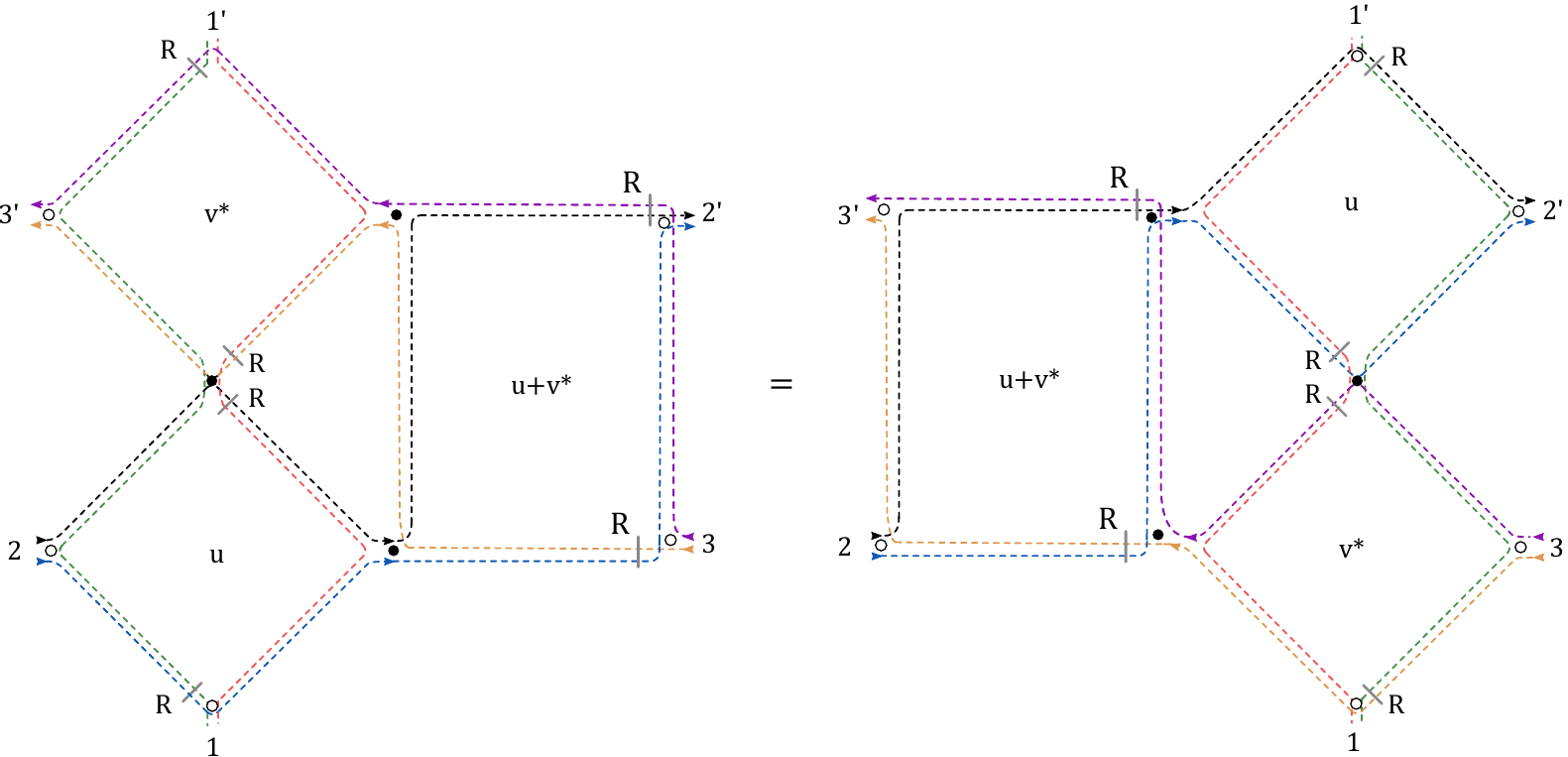}
\end{center}
\caption{Diagrammatic representation of the equation \eqref{YBE_tilda}. Different colors refer to different spinorial structures, black blobs indicate corresponding integrations and
circles indicate external coordinates without any integration.
Letters $u,v,u+v^*$ refer to the spectral parameters carried by the R-operators involved.}
\label{YBE_tilda_graph}
\end{figure}

The diagrammatic technique applied to the proof of the YBE can be repeated step-by-step to prove the relation $\rmat\rmat\rmat^{\dagger}=\rmat^{\dagger}\rmat\rmat$ \eqref{YBE_tilda} represented in Fig.\ref{YBE_tilda_graph}.
This kind of graphical computations are ubiquitous in our paper as they heavily simplify tedious analytic computations.
\noindent
In a similar way as the YBE, it is possible to prove that $\rmat$ and $\rmat^{\dagger}$ satisfy the simple algebra (see diagrammatic representation in Fig.\ref{YBE_tilda_graph})
\begin{equation}
\label{YBE_tilda}
\rmat_{12}(u){\rmat}_{32}(u+v^*)\rmat_{13}(v)^{\dagger}\,=\,\rmat_{13}(v)^{\dagger}{\rmat}_{32}(u+v^*)\rmat_{12}(u)\,.
\end{equation}
\\\\
\noindent

We introduce here for later purposes a factorization of the operator $\hat\rmat(u)$ as the product of
two operators $\hat \rmat_{12}(u) = \hat{\rmat}_{12}^{-}(u)\hat{\rmat}^{+}_{12}(u)$,
where
\begin{multline}
\label{R+}
\hat{\rmat}_{12}^{+}(u)\,=\,S_{\dot{\ell}_1\ell_1}\left(2-\Delta_1,x_{12}\right)\,
S_{\dot{\ell}_2\dot{\ell}_1}\left(u-\Delta_{-},\hat{p}_{1}\right)
\,S_{\ell_1\dot{\ell}_2}\left(u+\Delta_{+},x_{12}\right) =
\\
=\frac{[\mathbf{\overline{x}}_{12}]^{\dot{\ell}_1}
\mathbf{R}_{\dot{\ell}_1\ell_1}(2-\Delta_1)
[\mathbf{x}_{12}]^{\ell_1}}{x_{12}^{2(\Delta_1-2)}}\,
\frac{[\mathbf{\overline{p}}_{1}]^{\dot{\ell}_2}
\mathbf{R}_{\dot{\ell}_2\dot{\ell}_1}(u-\Delta_{-})
[\mathbf{p}_{1}]^{\dot{\ell}_1}}{\hat{p}_{1}^{2(\Delta_{-}-u)}}\,
\frac{[\mathbf{\overline{x}}_{12}]^{\ell_1}
\mathbf{R}_{\ell_1\dot{\ell}_2}(u+\Delta_{+})
[\mathbf{x}_{12}]^{\dot{\ell}_2}}{x_{12}^{-2(u+\Delta_{+})}}
\end{multline}
and
\begin{multline}
\label{R-}
\hat{\rmat}_{12}^{-}(u)\,=\,
S_{\dot{\ell}_1\ell_2}\left(u-\Delta_{+},x_{12}\right)\,
S_{\ell_2\ell_1}\left(u+\Delta_{-},\hat{p}_2\right)\,
S_{\ell_1\dot{\ell}_1}\left(\Delta_1-2,x_{12}\right) = \\
= \frac{[\mathbf{\overline{x}}_{12}]^{\dot{\ell}_1}
\mathbf{R}_{\dot{\ell}_1\ell_2}(u-\Delta_{+})
[\mathbf{x}_{12}]^{\ell_2}}{x_{12}^{2(\Delta_{+}-u)}}\,
\frac{[\mathbf{\overline{p}}_{2}]^{\ell_2}
\mathbf{R}_{\ell_2\ell_1}(u+\Delta_{-})
[\mathbf{p}_{2}]^{\ell_1}}{\hat{p}_{2}^{-2(u+\Delta_{-})}}\,
\frac{[\mathbf{\overline{x}}_{12}]^{\ell_1}
\mathbf{R}_{\ell_1\dot{\ell}_1}(\Delta_{1}-2)
[\mathbf{x}_{12}]^{\dot{\ell}_1}}{x_{12}^{2(2-\Delta_{1})}}\,.
\end{multline}
\begin{figure}[H]
\begin{center}
\includegraphics[scale=1.1]{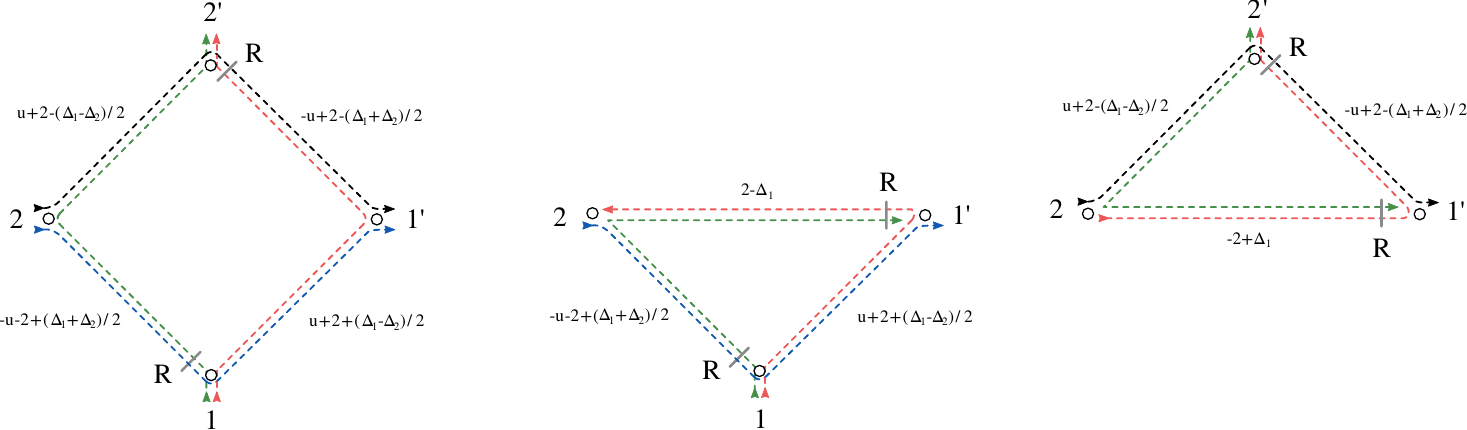}
\caption{\textbf{Left:} the kernel of the general $\hat \rmat$-operator in diagram formalism. It can be factorized into the product of two integral kernels $\hat \rmat_{\pm}$. \textbf{Center:} integral kernel of the operator $\hat \rmat_-(u)$ \textbf{Right:} integral kernel of the operator $\hat \rmat_+(u)$. When multiplied, the horizontal red/green lines in the two triangular kernels simplify, as their power is opposite as well as the argument of the two $\mathbf{R}$-matrices involved (see \eqref{unitar}) and the kernel of $\hat \rmat (u)$ is recovered.}
\end{center}
\label{R_fac}
\end{figure}

\section{Inhomogeneous spinning Fishnet}
\label{InFish}
The existence of an $\rmat$-operator acting on any two unitary irreps of the conformal group $SO(1,5)$ allows to define a family of integrable spin chains with conformal symmetry. These systems are the generalization of the $SU(2)$ Heisenberg magnet \cite{Faddeev:1996iy,Kulish:1981bi} to the conformal group $SO(1,5)$, being a chain of nearest-neighbour interacting sites. Each site carries a vector in the module of a given unitary irrep of the conformal group \cite{Tod:1977harm} - that is the wave function of a conformal particle in the Euclidean $4d$ space. The states of the spin chain with $L$ sites belong to the Hilbert space
\begin{equation}
\mathcal{V}=\mathbb{V}_1\otimes \mathbb{V}_2\otimes \cdots \otimes \mathbb{V}_L \,,
\end{equation}
where each $\mathbb{V}_k$ is the module of the representation $\left(\Delta_k,\ell_k,\dot{\ell}_k\right)$ - in general different at each site - and the scalar product on $\mathcal{V}$ is inherited by those on the sites $\mathbb{V}_k$ defined in \eqref{scalar_product}. The wave function of the system depends on the position of the particles $x^{\mu}_k\,\in\,\mathbb{R}$ and carries $\ell_k$- and $\dot{\ell}_k$-symmetric spinor indices
\begin{equation}
\Phi_{\mathbf{a_1\dot{a}_1\dots a_L\dot{a}_L}}(x_1,\dots,x_L)\,,\,\,\,\,\,\mathbf{a_k} = (a_{k,1},\dots,a_{k,\ell_k})\,,\,\,\,\mathbf{\dot a_k} = (\dot a_{k,1},\dots,\dot a_{k,\dot \ell_k})\,.
\end{equation}
%In this section we often encode the symmetric spinors as a polynomial in the auxiliary spinors $(\alpha_k,\beta_k)$ as explained in \eqref{}, and use the notation
%\begin{equation}
%\label{states}
%\Phi(x_1,\dots,x_L,\alpha_1,\dots,\alpha_L,\beta_1,\dots,\beta_L)=\langle{\beta}|\Phi(x_1,\dots,x_L)|{\alpha}\rangle\,.
%\end{equation}
Under the action of the conformal group these functions transform in the tensor product of the representations defined at each site. We consider a closed spin chain, where the $(L+1)$-th particle is identified with the $1$-st one, with periodic boundary conditions\footnote{This condition can be relaxed by the introduction of a twist at the boundary while preserving the integrability of the model.}. The dynamics of the system can be introduced starting from the definition of a \emph{monodromy matrix} operator
\begin{equation}
\label{monodromy_op}
\mathbb{T}_{1,\dots,L,a}(u)\,=\,\rmat_{1a}(u+\theta_1)\rmat_{2a}(u+\theta_2)\cdots \rmat_{La}(u+\theta_L)\,,
\end{equation}
which acts on the physical space of system $\mathcal{V}$ and on an auxiliary space $\mathbb{V}_a$, chosen to be the module of the unitary irrep $(\Delta_a,\ell_a,\dot{\ell}_a)$.
\begin{figure}[H]
\begin{center}
\includegraphics[scale=1]{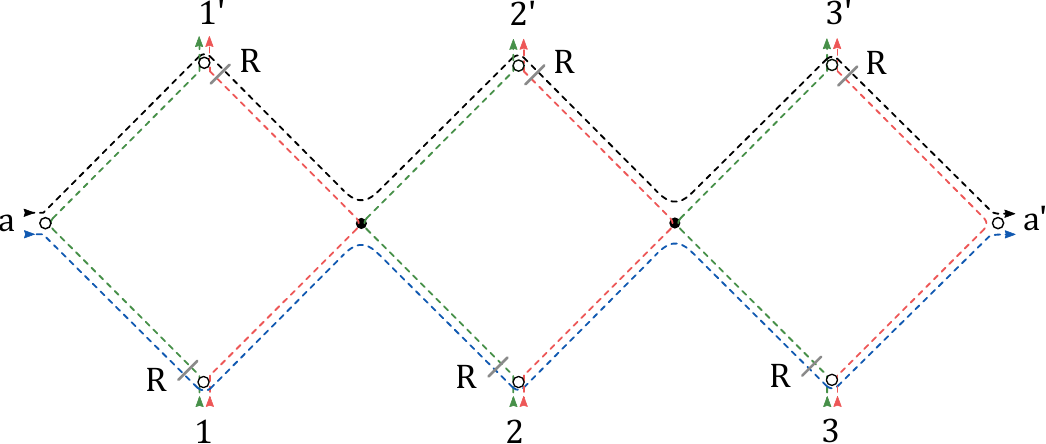}
\end{center}
\caption{Integral kernel of the monodromy matrix operator $\mathbb{T}_{1,2,3,a}(u)$ obtained from the convolution of operators $\rmat_{ka}(u+\theta_k)$ in the auxiliary space $\mathbb{V}_a$ for $k=1,2,3$. The black blobs are integrated, the circles are external coordinates. Black and blue lines denote the propagation of Weyl spinors in the auxiliary space $(\ell_a,0)$ and $(0,\dot{\ell}_a)$, while the red and green lines are Weyl spinor propagators in the representations $(\ell_k,0)$ or $(0,\dot{\ell}_k)$ of the physical spaces $\mathbb{V}_k$, $k=1,\dots,L$. The mixing of spinorial indices between auxiliary (black, blue) and physical (red, green) representations is given by the $SU(2)$ R-matrices denoted with a grey line. The transfer matrix $t^{(a)}(u)$ is obtained by the identification and integration of the points $x_a$ and $x_a'$.}
\end{figure}
\noindent
As a consequence of the YBE \eqref{YBE_inf} the monodromy operator satisfies the $RTT$ relation (Yangian algebra)
\begin{equation}
\label{RTT_Yang}
\rmat_{ab}(u-v) \mathbb{T}_{1,\dots,L,a}(u)\mathbb{T}_{1,\dots,L,b}(v)\,=\,\mathbb{T}_{1,\dots,L,a}(u) \, \mathbb{T}_{1,\dots,L,b}(v)\, \rmat_{ab}(u-v)\,.
\end{equation}
The algebra \eqref{RTT_Yang} allows to define an infinite family of commuting  \emph{transfer matrix} operators
\begin{equation}
\label{t_comm}
t^{(a)}(u) t^{(b)}(v) =  t^{(b)}(v) t^{(a)}(u)\,,
\end{equation}
as the infinite dimensional trace in the auxiliary spaces $\mathbb{V}_a\otimes \mathbb{V}_b$ of monodromies
\begin{equation}
\label{t_matrix_op}
t^{(a)}(u)\equiv t^{(a)}_{1,\dots,L}(u)\,\equiv \,\text{Tr}_{\mathbb{V}_a} \left(\mathbb{T}_{1,\dots,L,a}(u)\right)\,,
\end{equation}
Each transfer matrix depends through the label $a$ on one of the infinite possible principal series irreps chosen for the auxiliary space.
As analytic functions of the parameter $u\,\in\,\mathbb{R}$, the operators $t^{(a)}(u)$ can be Taylor expanded around a chosen $u=u_0$ leading to an infinite tower of commuting operators\begin{equation}
H^{(a)}_k = \left[\frac{1}{k!}\frac{d^k}{du^k} t^{(a)}(u)\right]_{u=u_0}\,,\,\,\,\, H^{(a)}_i H^{(b)}_j\,=\,H^{(b)}_j H^{(a)}_i\,.
\end{equation}
The definition of an integrable model with commuting charges $H^{(a)}$ requires that such operators are diagonalizable.
In other words, we should require the operator $t^{(a)}(u)$ to be normal, i.e. to commute with its hermitean conjugate\begin{equation}
\left[t^{(a)}(u),t^{(a)}(u)^{\dagger}\right]\,=\,0\,.
\end{equation}
The previous equation is actually a particular case of the general property
\begin{equation}
\label{t_normal}
t^{(a)}(u) \,(t^{(b)}(v))^{\dagger}\,=\,(t^{(b)}(v))^{\dagger} \,t^{(a)}(u)\,,\,\,\,\,\,\forall\,u,v\,,
\end{equation}
valid for any two auxiliary space representations $(\Delta_a,\ell_a,\dot{\ell}_a)$ and $(\Delta_b,\ell_b,\dot{\ell}_b)$ under the constraint
\begin{equation}
\label{inhom_constr}
\theta_k+\theta_k^* \,=\,\theta_j+\theta_j^*\,,\,\,\,\,\,\forall j,k\,=\,1,\dots,L\,.
\end{equation}
The proof of \eqref{t_normal} makes use of the R-operator property \eqref{YBE_tilda}, in the same way that \eqref{t_comm} follows from the YBE. First, we can write the hermitean conjugate of the monodromy operator
\begin{equation}
\mathbb{T}_{1,\dots,L,a}^{\dagger}(u)\,=\,\rmat_{La}(u+\theta_L)^{\dagger}\cdots \rmat_{2a}(u+\theta_2)^{\dagger}\rmat_{1a}(u+\theta_1)^{\dagger}\,,
\end{equation}
where by $\dagger$ we mean hermitean conjugation in both spaces, physical and auxiliary, for each R-operator. Making use of \eqref{YBE_tilda} we can write a $T R T^{\dagger}$ relation
\begin{equation}
\mathbb{T}_{1,\dots,L,a}(u){\mathcal{R}}_{ba}(u+v^*+\theta+\theta^*)\,\mathbb{T}_{1,\dots,L,b}(v)^{\dagger}\,=\,\mathbb{T}_{1,\dots,L,b}(v)^{\dagger}\,{\mathcal{R}}_{ba}(u+v^*+\theta+\theta^*)\,\mathbb{T}_{1,\dots,L,a}(u)\,,
\end{equation}
where $\theta+\theta^*=\theta_k+\theta_k^*$ for any $k=1,\dots,L$.
The algebra \eqref{t_normal} follows from the trace over auxiliary spaces in the previous equation, where
\begin{equation}
(t^{(a)}(u))^{\dagger}\,=\,\text{Tr}_{\mathbb{V}_a} \left(\mathbb{T}_{1,\dots,L,a}^{\dagger}(u)\right)\,=\,\text{Tr}_{\mathbb{V}_a} \left(\rmat_{La}(u+\theta_L)^{\dagger}\cdots \rmat_{2a}(u+\theta_2)^{\dagger}\rmat_{1a}(u+\theta_1)^{\dagger}\right)\,.
\end{equation}
Notice that the constraint \eqref{inhom_constr} is compatible with the choice
\begin{equation}
\theta_k= -\frac{\Delta_k}{2} = -1 -i\nu_k\,,
\end{equation}
which is particularly suitable to reduce $t^{(a)}(u)$ at special values of $u$ as explained in the next section.
%\begin{align}
%\label{t_pm_comm_dag}
%begin{aligned}
%&[t^{(a)}_{\pm}(u),(t^{(b)}_{\pm}(v))^{\dagger}]\,=\,0\,,\\
%&[t^{(a)}_{\pm}(u),(t^{(b)}_{\mp}(v))^{\dagger}]\,=\,0\,.
%\end{aligned}
%\end{align}
%The proofs of \eqref{t_normal} are very similar to the one of \eqref{t_pm_comm} and are achieved by the same passages, therefore will not be repeated here. The only difference, lies in the type of exchange relation due to the different position of the $SU(2)$ $\mathbf{R}$-matrices in the kernels.

%\begin{equation}
%\tilde{\mathcal{R}}_{ab}(u+v)\rmat_{1a}(u)\rmat_{1b}(v)^{\dagger}%\,=\,\rmat_{1b}(v)^{\dagger}\rmat_{1a}(u)\tilde{\mathcal{R}}_{ab}(u+v)
%\end{equation}
Whenever \eqref{inhom_constr} is satisfied the operators $t^{(a)}(u)$ are normal operators for any choice of auxiliary space and $u$, therefore for its Taylor coefficients it holds $[H_k^{(a)},H_j^{(b)\,\dagger}]=0$. Thus $H_k^{(a)}$ are a continuous infinity of mutually diagonalizable operators. The Hamiltonian of the spin chain is usually picked to be the logarithmic derivative of the transfer matrix \cite{Tarasov:1983cj}
\begin{equation}
\label{NN_hamilton}
\mathbb{H}\,\equiv\, \left(H_0^{(a)}\right)^{-1}\,H_1^{(a)}(u_0)\,.
\end{equation}
The operator \eqref{NN_hamilton} takes the form of a sum of two-particle Hamiltonians acting on nearest-neighbouring particles, prior to the sufficient condition that at $u=u_0$ each R-operator $\rmat_{ia}(u)$ inside the trace in $t^{(a)}(u)$ reduces to the permutation $\mathbb{P}_{ia}$. This fact is possible only for the homogeneous chain, that is when the representations of the conformal group in each site and in auxiliary space
are the same $(\Delta_k,\ell_k,\dot{\ell}_k) =(\Delta_a,\ell_a,\dot{\ell}_a) = (\Delta,\ell,\dot{\ell})$, and all shifts are the same $\theta_k=\theta$.
For convenience we set $\theta=0$, so that $u_0=0$. The explicit expression for the two-particle Hamiltonian can be obtained from the solution of the braid equation~\eqref{braid_sol}; in the case of homogeneous chain we have
$\Delta_- =0\,,\Delta_+ = \Delta-2$ and $\ell_1=\ell_2=\ell\,,\dot{\ell}_1=\dot{\ell}_2=\dot{\ell}$
so that
\begin{align}\label{RH}
\hat\rmat_{12}(u) = \frac{[\mathbf{\overline{x}}_{12}]^{\dot{\ell}}
\mathbf{R}_{\dot{\ell}\ell}(u-\Delta_{+})
[\mathbf{x}_{12}]^{\ell}}{x_{12}^{2\left(-u+\Delta_{+}\right)}}\,
\frac{[\mathbf{\overline{p}}_{2}]^{\ell}
\mathbf{R}_{\ell\ell}(u)
[\mathbf{p}_{2}]^{\ell}}
{\hat{p}_{2}^{-2u}}\,
\frac{[\mathbf{\overline{p}}_{1}]^{\dot{\ell}}
\mathbf{R}_{\dot{\ell}\dot{\ell}}(u)
[\mathbf{p}_{1}]^{\dot{\ell}}}
{\hat{p}_{1}^{-2u}}\,
\frac{[\mathbf{\overline{x}}_{12}]^{\ell}\mathbf{R}_{\ell\dot{\ell}}(u+\Delta_{+})
[\mathbf{x}_{12}]^{\dot\ell}}
{x_{12}^{2\left(-u-\Delta_{+}\right)}}\,.
\end{align}
The condition $\hat\rmat_{12}(0)=\mathbb{P}_{12}\rmat_{12}(0) = \mathbbm{1}$ is satisfied and
\begin{align}
\label{H12_spin}
\mathbb{H}_{12} = \hat\rmat^{\prime}_{12}(u) = 2\ln x_{12}^{2}
+ S_{12}^{-1}\,\left(\ln \hat{p}_1^2+\ln \hat{p}_2^2+
[\mathbf{\overline{p}}_{1}]^{\dot{\ell}}
\mathbf{R}^{\prime}_{\dot{\ell}\dot{\ell}}(0)
[\mathbf{p}_{1}]^{\dot{\ell}}+
[\mathbf{\overline{p}}_{2}]^{\ell}
\mathbf{R}^{\prime}_{\ell\ell}(0)
[\mathbf{p}_{2}]^{\ell}
\right)\,S_{12}\,,
\end{align}
where
\begin{align*}
S_{12} = x_{12}^{2\Delta_{+}}[\mathbf{\overline{x}}_{12}]^{\ell}
\mathbf{R}_{\ell\dot{\ell}}(\Delta_{+})
[\mathbf{x}_{12}]^{\dot{\ell}}\,.
\end{align*}
Note that operators $\mathbf{R}^{\prime}_{\dot{\ell}\dot{\ell}}(0)$ and $\mathbf{R}^{\prime}_{\ell\ell}(0)$ are the standard Hamiltonians~\cite{Tarasov:1983cj}
of the $SU(2)$-invariant spin chains acting  for the symmetric representations $\dot{\ell}$ and $\ell$ respectively. In the case of zero spins $\mathbb{H}_{12}$ reduces to the $4d$ version of Lipatov's Hamiltonian~\cite{Chicherin:2012yn}
\begin{align}
\mathbb{H}_{12} = \hat\rmat^{\prime}_{12}(u) = 2\ln x_{12}^{2}
+ x_{12}^{-2\Delta_{+}}\,\left(\ln \hat{p}_1^2+\ln \hat{p}_2^2\right)\,
x_{12}^{2\Delta_{+}}\,,
\end{align}
and \eqref{H12_spin} is its spinning generalization.

It follows from the definition of $\mathbb{H}$, that any other operator $H^{(a)}_k$ is a conserved charge\footnote{In this respect, any other $H^{(a)}_k$ or hermitean linear combinations thereof can be taken as the Hamiltonian of the system. This includes infinite linear combinations, for instance $H_0^{(a)} = t^{(a)}(u_0)$ itself.} \begin{equation}
\left[\mathbb{H}, H^{(a)}_i\right]\,=\, 0\,.
\end{equation}
The number of degrees of freedom of the system of $L$ particles is $6L$, since each particle in the chain has two $SU(2)$ spins and four coordinates, i.e. six degrees of freedom. One can extract from the infinite tower of conserved charges $\{H^{(a)}_k\}$, a maximal subset of $6L$ linearly independent operators so to establish the complete integrability of the quantum model.

\begin{figure}[H]
\begin{center}
\includegraphics[scale=0.75]{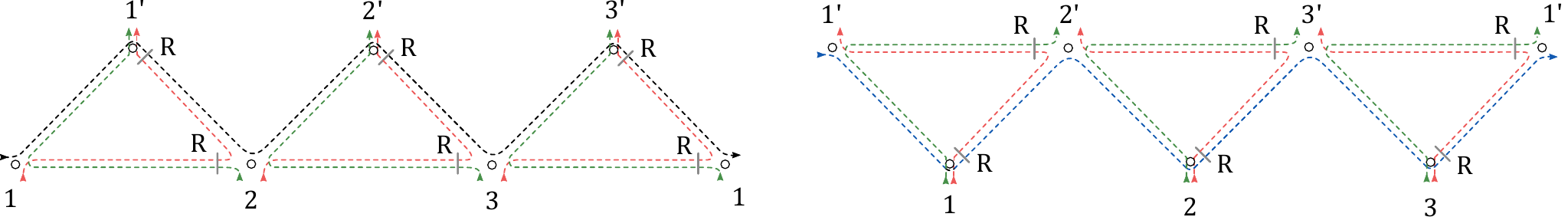}
\end{center}
\caption{\textbf{Left:} integral kernel $t_{+}(x_1,x_2,x_3|x_1',x_2',x_3')(u)$ of the operator $t_{+}(u)$. The black dashed segments transform in the $\ell_a$-symmetric irrep of $SU(2)$. \textbf{Right:} integral kernel $t_{-}(x_1,x_2,x_3|x_1',x_2',x_3')(u)$ of the operator $t_{-}(u)$. The blue dashed segments transform in the $\dot{\ell}_a$-symmetric irrep of $SU(2)$.}
\label{t_pm_diag}
\end{figure}

An alternative, equivalent, definition of transfer matrices can be done starting from the factorization \eqref{R+} \eqref{R-}, inherited by the transfer matrix operator
\begin{align}
\label{t_pm}
t^{(a)}(u) = t^{(a)}_-(u)\,t^{(a)}_+(u)\,.
\end{align}
%\begin{align}
%\begin{aligned}
%&{\rmat}^{(a)+}_{12}(u) = \int d^4 x_a \,\rmat^{+}_{1a}(u) \delta^{(4)}(x_a-x_2)\,,\\&{\rmat}^{(a)-}_{12}(u) = \int d^4 x_a \,\delta^{(4)}(x_a-x_1)\mathbb{P}_{1a}\rmat^{-}_{1a}(u) \,.
%\end{aligned}
%\end{align}
The interpretation of \eqref{t_pm} becomes very transparent in the diagrammatic notation of Fig.\ref{t_pm_diag}.
Both operators $t^{(a)}_{\pm}(u)$ act on the physical space $\mathcal{V}$ and carry respectively only one spin $(\ell_a,0)$ or $(0,\dot{\ell}_a)$ in the auxiliary space, as made clear in Fig.\ref{t_pm_diag}.
By means of the interchange relations of Fig.\ref{Inter_I} one can directly verify the commutation
\begin{equation}
\label{t_pm_comm}
t^{(a)}_{+}(u)t^{(b)}_-(v)=t^{(b)}_-(v)t^{(a)}_+(u)\,.
\end{equation}
Indeed, starting from the l.h.s. of \eqref{t_pm_comm}, depicted on the left, it is possible to insert the identity along the convolution in the auxiliary space in the form $\hat S_{ab}(u-v,x_{11'})\hat S_{ab}(v-u,x_{11'})$, as on the right picture.
\begin{figure}[H]
\begin{center}
\includegraphics[scale=0.6]{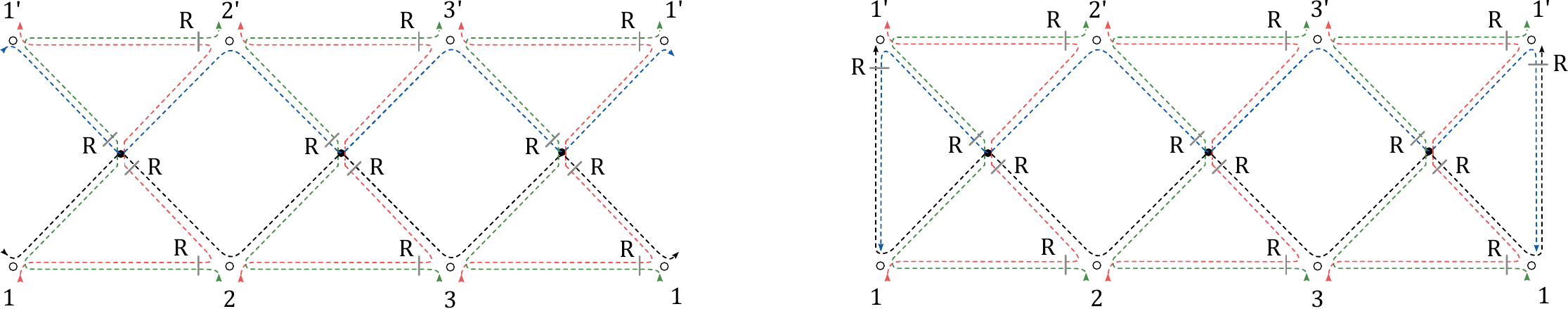}
\end{center}
\end{figure}
The interchange relation of Fig.\ref{Inter_I} can be applied in order to move the black/blue lines $\hat S_{ab}(u-v,x_{11'})$ from the left to the right of the diagram,  moving downstairs the horizontal red/green lines in the upper part of the diagram. One can easily recognize that the result amounts to the r.h.s of \eqref{t_pm_comm}.
\begin{figure}[H]
\begin{center}
\includegraphics[scale=0.75]{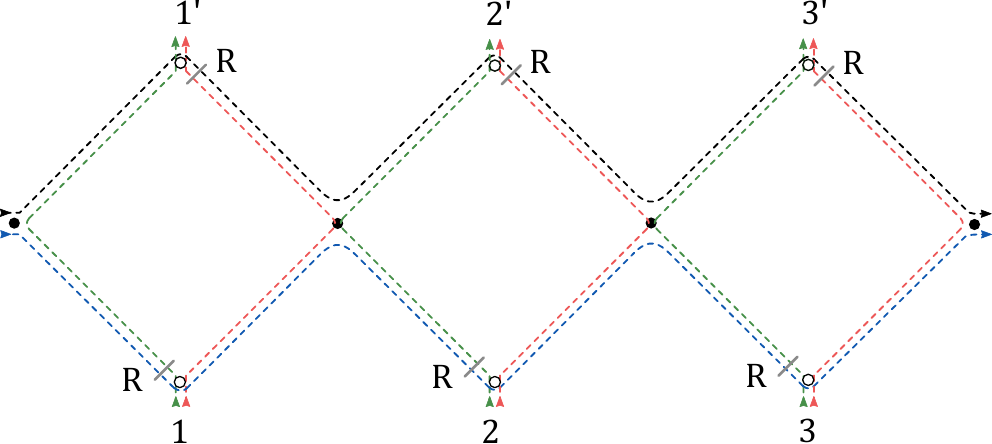}
\end{center}
\end{figure}
\noindent
Moreover, under the constraint \eqref{inhom_constr}, the following commutation holds
\begin{equation}
\label{t_pp_comm}
t^{(a)}_{\pm}(u)t^{(b)}_{\pm}(v)=t^{(b)}_{\pm}(v)t^{(a)}_{\pm}(u)\,.
\end{equation}
The proof can be delivered via diagrams in few steps: starting from $t^{(a)}_{+}(u)t^{(b)}_{+}(v)$ - depicted on the left - the triangles corresponding to $\rmat^{+}(v)$ get opened into star integrals by means of \eqref{STR_good}, obtaining the picture on the right:
\begin{figure}[H]
\begin{center}
\includegraphics[scale=0.6]{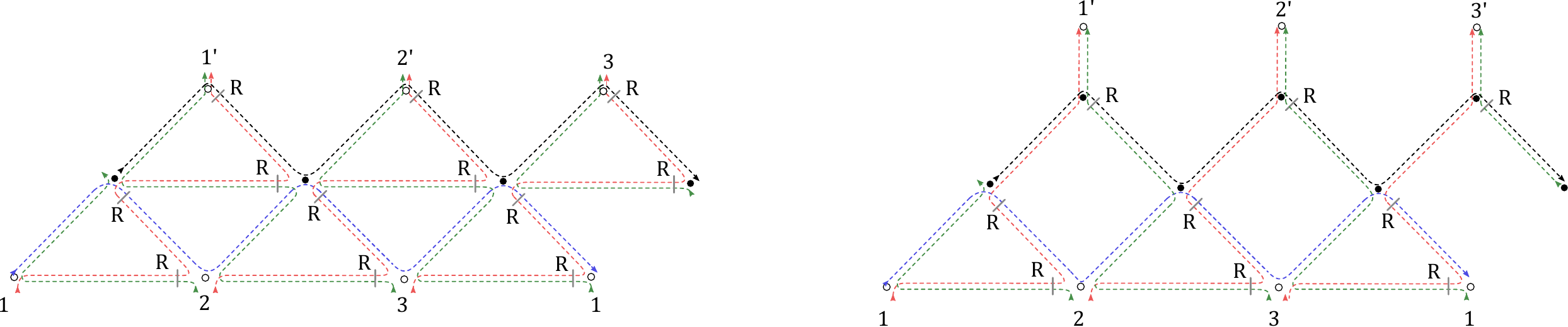}
\end{center}
\end{figure}
\noindent
The equation \eqref{t_pp_comm} amounts to the exchange of auxiliary spaces $\mathbb{V}_a$ and $\mathbb{V}_b$ and the spectral parameters. In order to do so we introduce along the auxiliary spaces (blue/black dashed lines) the identity in the form $S_{ab}(u-v)S_{ab}(v-u)=\mathbbm{1}$, that is two couples of black/blue vertical lines, as in the left picture. Now, by means of the interchange relation of Fig.\ref{Inter_I} one can move the lines on the left, namely $S_{ab}(u-v)$, to the right until they collide with $S_{ba}(v-a)$ and cancel due to \eqref{reflect2}. The overall effect is an exchange of blue/black dashed lines as well as of the parameters $u$ and $v$ (right picture).
\begin{figure}[H]
\begin{center}
\includegraphics[scale=0.6]{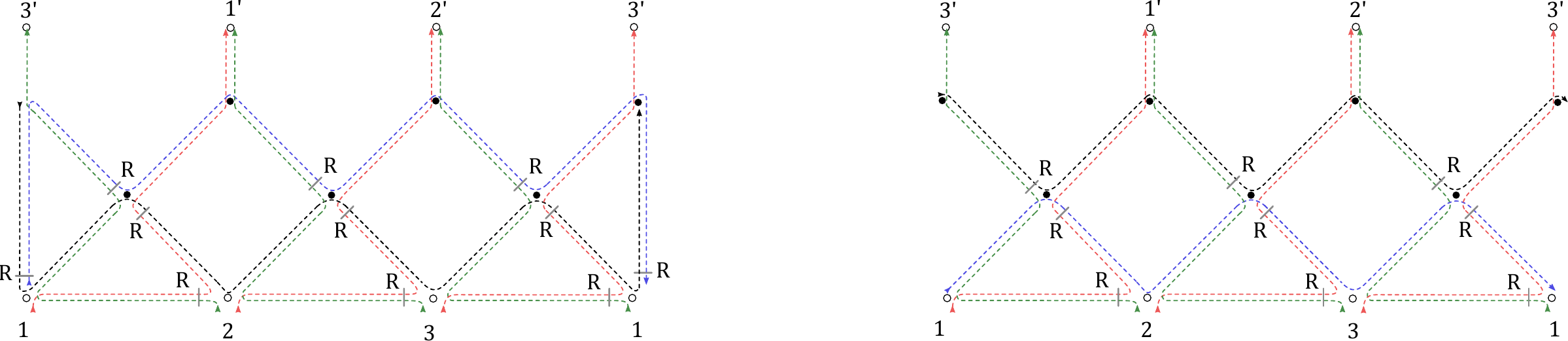}
\end{center}
\end{figure}
Finally, it is enough to perform the inverse of the first step, integrating the star integrals into triangles $\rmat^{+}(u)$, recovering the r.h.s. of \eqref{t_pp_comm} $\square$.
%The factorization \eqref{t_pm} is remarkable as the two factors are related by hermitean conjugation, as it follows from \eqref{}, namely
%\begin{equation}
%t^{(a)}_{+}(u)^{\dagger}=t^{(a)}_{-}(u)
%\end{equation}

In order to prove the integrability of the model, we finally observe that the property \eqref{R_hc} and the definition \eqref{t_pm} imply that the hermitean conjugate of $t_{+}^{(a)}(u)$ is an integral operator with the same kernel of $t_{-}^{(a)}(u^*)$, apart from the flow of auxiliary space $\boldsymbol{\sigma}$, $ \boldsymbol{\overline{\sigma}}$ and the relative position of $\mathbf{R}(u)$ matrices which is the opposite. In diagrammatic form this amounts to
\begin{figure}[H]
\begin{center}
\includegraphics[scale=0.75]{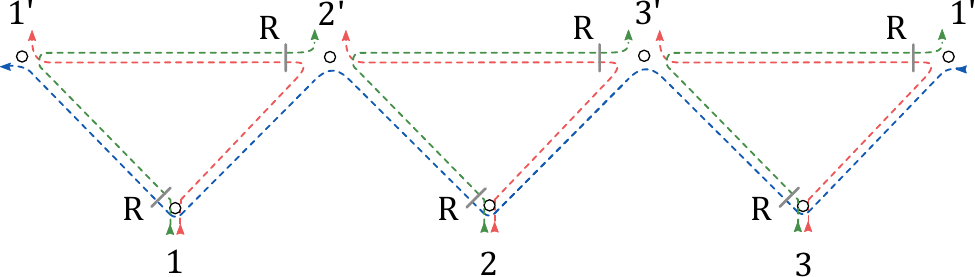}
\end{center}
\end{figure}
\noindent
and from this observation it follows that the proof of \eqref{t_pm_diag} can be repeated step-by-step in order to proof that
\begin{equation}
\label{t_pm_normal}
t_{\pm}^{(a)}(u)t_{\pm}^{(b)}(v)^{\dagger}=t_{\pm}^{(b)}(v)^{\dagger}t_{\pm}^{(a)}(u)\,,
\end{equation}
with the only difference that the interchange relation involved instead of Fig.\ref{Inter_I} is the one of Fig.\ref{Inter_II}.
The equations \eqref{t_pp_comm} and \eqref{t_pm_normal} show that the ``half" transfer matrices $t^{(a)}_{\pm}(u)$ are an alternative way to define the same integrable model, as their Taylor expansion in $u$ generates two families of conserved charges.
\subsection{Integrable correlators in conformal field theories}
\label{subsect:fish}
\begin{figure}[H]
\begin{center}
\includegraphics[scale=0.60]{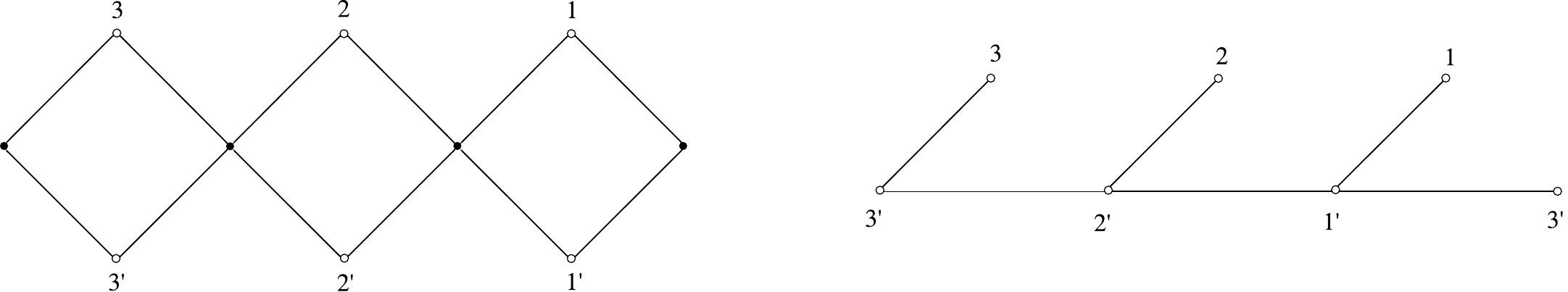}
\caption{\textbf{Left:} Integral kernel of the transfer matrix operator $t^{(a)}(u)$ for a spin chain of length $L=3$, for auxiliary and physical spaces in the representation of zero spins $\ell_a=\ell_k=0$ and $\dot{\ell}_a=\dot{\ell}_k=0$ and $\Delta_k=\Delta_a=1$. \textbf{Right:} at the point $u=-1$ and $\theta=0$ (or, equivalently, $u\to -1/2$ and $\theta=-\Delta/2=-1/2$) the transfer matrix kernel reduces to the graph-building operator $\hat B$ for the planar Feynman integrals of the bi-scalar Fishnet theory \eqref{Fishnet_L}.}
\end{center}
\label{builder_fish}
\end{figure}

\begin{figure}[H]
\begin{center}
\includegraphics[scale=0.55]{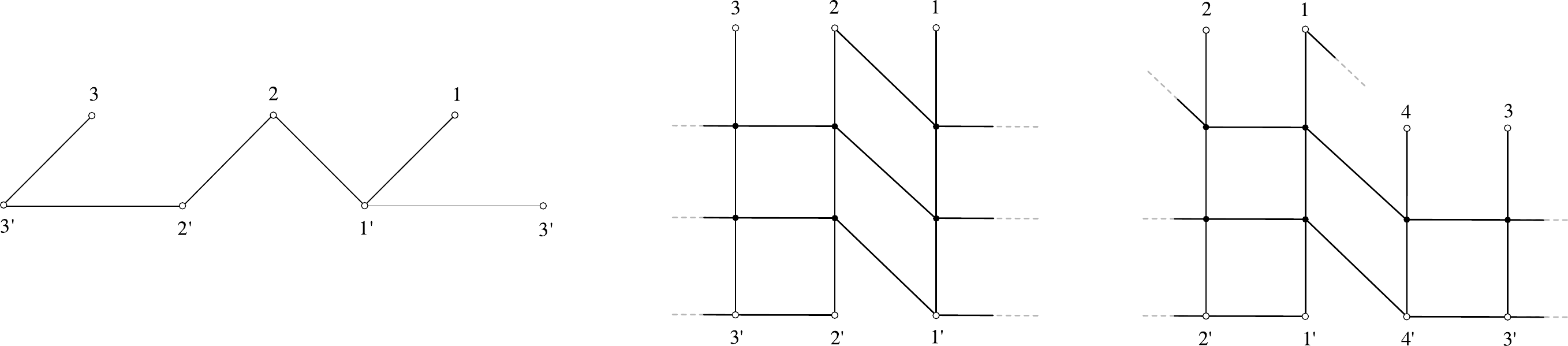}
\caption{\textbf{Left:} Reduction at the point $u\to-1/2$ of the kernel of the transfer matrix operator $t^{(a)}(u)$ for a spin chain of length $L=3$, for auxiliary and physical spaces in the representation of zero spins $\ell_a=\ell_k=0$, $\dot{\ell}_a=\dot{\ell}_k=0$ and scaling dimensions $\Delta_{1,3}=\Delta_a=1$ and $\Delta_2=2$. \textbf{Center:} the iterative composition of the reduced transfer matrix is the graph-builder of the bulk of Feynman integrals contributing to the two point function of $\text{Tr}\left[\phi_1^2\phi_2\right]$ \textbf{Right:} a possible choice of boundary condition for the graphs of  $\text{Tr}\left[\phi_1^3\phi_2\right]$, corresponding to the term $k=2$ in \eqref{fishnet_magnon}. The first operator from the top of the graph is $\hat B'_2$ while the second and third are $\hat B'_4$ and add both a wrapping around the graph.}
\label{build_fish_magg}
\end{center}
\end{figure}

In the papers \cite{Gurdogan:2015csr,Gromov:2017cja,Gromov:2018hut}, the authors dealt with the homogeneous chain of particles in the scalar representation $(\Delta_k,\ell_k,\dot\ell_k)=(\Delta,0,0)$ and for an auxiliary space in the same representation. It was observed that for $\Delta=1$ at the point $u=-1$  and $\theta=0$ the transfer matrix operator $t^{(a)}(u)$ reduces to the \emph{graph-building} operator for a class of planar Feynman diagrams with regular square-lattice topology (see Fig.\ref{builder_fish}), and its spectral problem was directly related to the spectra of scaling dimensions of the single-trace \emph{vacua} $\text{Tr}\left[\phi_1^L\right](x)$ of the conformal Fishnet theory (FCFT) introduced in \cite{Gurdogan:2015csr}
\begin{equation}
\label{Fishnet_L}
 {\cal L}_{\phi}= \frac{N}{2}\Tr
    \left(\p^\mu\phi^\dagger_1 \p_\mu\phi^1+\p^\mu\phi^\dagger_2 \p_\mu\phi^2+2\xi^2\,\phi_1^\dagger \phi_2^\dagger \phi^1\phi^2\right)\,,
    \end{equation}
    where $\phi_k$ are $N\times N$ complex scalar fields in the adjoint representation of $SU(N$).
In fact, the scaling behaviour of the two point functions of vacua at $L>2$ at finite coupling is captured by a geometric series
\begin{equation}
\label{fishnet_integr}
\lim_{y\to \infty}\langle \text{Tr}\left[\phi_1^L\right](x)\text{Tr}\left[\phi_1^L\right]^{\dagger}(y)\rangle \propto \big \langle x| \frac{1}{1-\xi^{2L} \hat B} | 1 \rangle\,,\,\,\,\,\, |x\rangle = \prod_{k=1}^L \delta^{(4)}(x_k-x)\,,\,\,|1\rangle =1\,,
\end{equation}
where $\hat B$ is the operator defined by the limit $u \to -1$ of the transfer matrix. The Feynman integrals contributing to the correlator in perturbation theory have a simple square-lattice \emph{fishnet} topology  and they are built by iterative composition of the operator $\xi^{2L}\hat B$, which therefore plays the role of integral Bethe-Salpeter kernel of the correlator, and the geometric series follows from the re-summation of all perturbative orders. The equation \eqref{fishnet_integr} is a statement of the quantum integrability of the FCTF vacua, as it maps the spectral problem for the scaling dimensions to an integrable quantum chain of particles.
The above analysis can be extended to other operators than the vacua, for instance the ones obtained by insertion of an \emph{excitation} $\phi_2$ upon the vacuum of $L>1$ fields $\phi_1$
\begin{equation}
\mathcal{O}'_{L}(x)=\text{Tr}\left[\phi_1^L \phi_2 \right](x)\,,
\end{equation}
Its two-point function has a scaling behaviour captured by the expression
\begin{equation}
\label{fishnet_magnon}
\lim_{y\to \infty} \, \langle \mathcal{O}'_{L}(x) \mathcal{O}'_{L}(y)^{\dagger}\rangle \propto \sum_{k=0}^{L-1} \frac{\xi^{2k}}{(4\pi^2)^{2k}} \big \langle x| {\hat B'_{k}} \frac{1}{1-\xi^{2L}  \hat B'_L} | 1 \rangle\,,
\end{equation}
where here the graph-builder $\hat B'_{L}$ is the limit $u\to-1/2$ of the transfer matrix of a spinless inhomogeneous chain of $L$ particles, out of which $L-1$ have dimension $\Delta=1$ ($\theta=-1/2$) and stand for a field $\phi_1$, and one has dimension $\Delta=2$ ($\theta=-1$) and stands for the pair of fields $\phi_1\phi_2$ (see Fig.\ref{build_fish_magg}). The sum over $k$ in \eqref{fishnet_magnon} involves the operator $\hat B'_k$ acting on the first $k$ sites starting from the ``magnonic one" $\Delta=2$, and accounts for the different boundary conditions of the graphs, while the geometric series in $\hat B'_L$ captures the graphs bulk. The class of Feynman integrals entering the perturbative expansion of such correlators is depicted in Fig.\ref{build_fish_magg} together with the graph-building kernel, so to make clear the origin of the geometric series and boundary terms in \eqref{fishnet_magnon}.
The arguments of the previous paragraphs can be repeated in full generality for every local operator of the bi-scalar theory
\eqref{Fishnet_L}, by an appropriate choice of the representations in the physical spaces of the chain and of inhomogeinities parameters \footnote{For a complete analysis of how to realize FCTF two-point functions starting from a spin-less conformal spin chain with inhomogeinities the reader can refer to \cite{Gromov2020}}.
Similarly, one can realize via a conformal chain any other correlator of the theory, since the bulk of Feynman integrals is a square-lattice and the only leftover freedom is the choice of boundary conditions for the quantum magnet - imposed by the choice of the correlator (see for example the four-point correlators \cite{Basso:2017jwq,Derkachov2020} corresponding to an open conformal chain, or $n$-point correlators analyzed in \cite{Chicherin:2017frs,Chicherin:2017cns}).

\begin{figure}
\begin{center}
\includegraphics[scale=1.28]{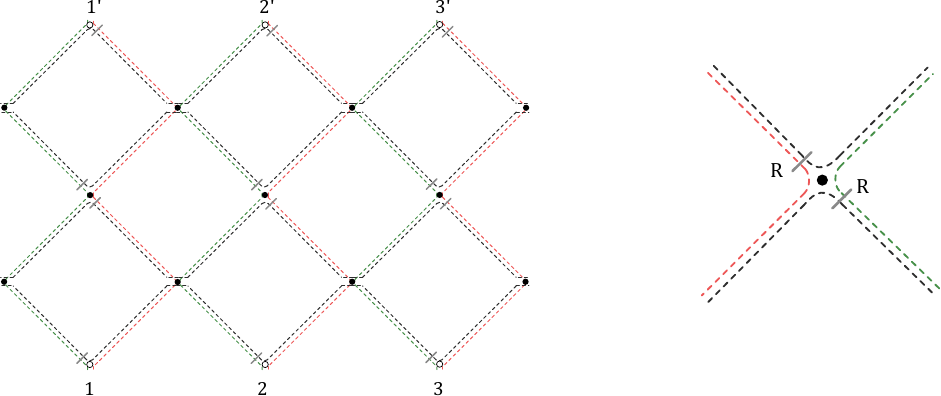}
\end{center}
\caption{\textbf{Left:} Spinning inhomogeneous fishnet Feynman diagram: any red line and green line carries a different Weyl spinor representation $(\Delta_k,\ell_k,0)$, $(\Delta_k,0,\dot{\ell}_k)$, while black lines are associated to the auxiliary space representations and their representation is set to $(\Delta_a,\ell_a,0)$ and $(\Delta_a,0,\dot{\ell}_a)$. Each integration point on the left is identified with the corresponding on the right, so the graph is wrapped on a cylinder. \textbf{Right:} quartic vertex of the planar lattice.}
\label{Spinning_fish}
\end{figure}

In the previous section we have showed that the transfer matrices $t^{(a)}(u)$ can be worked out in the language of Feynman integrals for any unitary irreps $\mathbb{V}_i$ and $\mathbb{V}_a$ of the conformal group. Therefore, $t^{(a)}(u)$ can be regarded as the \emph{graph-building} operator for a class of planar Feynman diagrams with square lattice topology, which we can dub \emph{spinning} (for some $(\ell_k,\dot{\ell_k})\neq(0,0))$ and \emph{inhomogeneous} (for $\mathbb{V}_k\neq \mathbb{V}_j$) fishnets, represented in Fig.\ref{Spinning_fish}. The quartic vertex of the lattice is scale invariant and mixes the $SU(2)$ spinor indices of the incoming/outcoming fields by means of the fused $\mathbf{R}$-matrix \eqref{RR_fused}.

The graph-building operators $t^{(a)}(u)$ of the spinning inhomogeneous fishnet for an appropriate choice of physical and auxiliary space representations describe other correlators in the broader context of chiral (a.k.a. generalized fishnet) CFTs, proving such correlators to be integrable from a spin chain picture. For example, an homogeneous spinning fishnet of Weyl fermions describes the strong deformation limit defined in \cite{Pittelli2019} of the $\mathcal{N}=2$ SYM planar two point correlators among the \emph{fermionic vacua} \begin{equation}
\mathcal{O}_{a_1\dots a_L}(x)=\text{Tr}\left[(\lambda_1)_{a_1}\cdots(\lambda_1)_{a_L} \right](x)\,,\,\,\, a_j=1,2\,.
\end{equation}
In this case the Feynman integrals of the perturbative expansion are described by an hexagonal lattice of Yukawa vertices, where fermions propagate along between the two operators (physical space), and scalar fields form wrappings in the orthogonal direction (auxiliary space), therefore the spin chain is defined by $(\Delta_k,\ell_k,\dot{\ell}_k)=(3/2,0,1)$ and $(\Delta_a,\ell_a,\dot{\ell}_a)=(1,0,0)$.
\begin{figure}[H]
\includegraphics[scale=0.50]{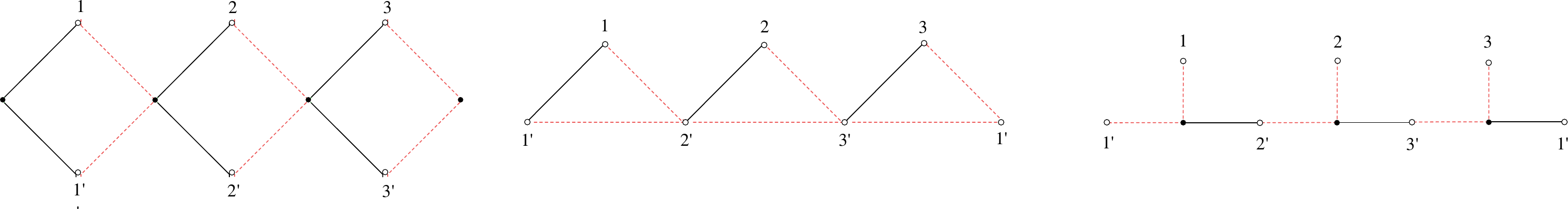}
\caption{\textbf{Left:} Integral kernel of the transfer matrix operator $t^{(a)}(u)$ for a spin chain of length $L=3$, for spinless auxiliary space $\ell_a=\dot{\ell}_a=0$ and physical spaces in the Weyl fermion representation $(3/2,0,1)$. \textbf{Center:} the transfer matrix kernel at $u \to -5/4$ and $\theta=0$ takes the shape of a triangular lattice wrapped onto a cylinder. \textbf{Right:} by means of star-triangle duality the triangular lattice can be re-written as an hexagonal lattice of Yukawa vertices, where fermions propagate along the cylinder and scalar around it.}
\end{figure}
Another class of integrable correlators described by honeycomb Feynman integrals - though with different boundary conditions - are the \emph{bosonic vacua} $\text{Tr}\left[\phi^L\right](x)$ in the same planar theory. In this case the fermions propagate around forming wrappings (i.e. auxiliary space), and scalar fields propagating between the two operators (physical space). Therefore one deals - as for the FCFT \eqref{Fishnet_L} - with the scalar homogeneous model $(\Delta_k,\ell_k,\dot{\ell}_k)=(1,0,0)$, but the graph-builder is a transfer matrix with fermionic auxiliary space $(\Delta_a,\ell_a,\dot{\ell}_a)=(3/2,0,1)$.
\begin{figure}[H]
\includegraphics[scale=0.51]{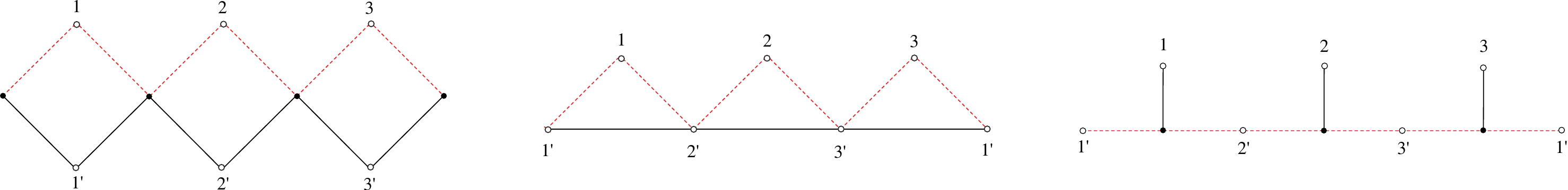}
\caption{\textbf{Left:} Integral kernel of the transfer matrix operator $t^{(a)}(u)$ for a spin chain of length $L=3$, for auxiliary space in the Weyl fermion representation $(3/2,0,1)$ and physical spaces in the the scalar representation $(1,0,0)$. \textbf{Center:} the transfer matrix kernel at $u=-5/4$ takes the shape of a triangular lattice wrapped onto a cylinder. \textbf{Right:} by means of star-triangle duality the triangular lattice can be re-written as an hexagonal lattice, with fermions propagating in angular direction and scalars along the cylinder.}
\end{figure}
Furthermore, the extension to inhomogeneous chain allows to mix scalars $(1,0,0)$ and Weyl fermions $(3/2,0,1)$ in the physical space, describing new classes of planar Feynman integrals. The transfer matrix of such a chain for $L$ scalar particles and $N$ fermionic ones is the graph builder of Feynman diagrams that mix the topology of square and \emph{honeycomb} lattice. Such a class of integrals describes completely the perturbation theory of two-point planar correlators of a single-trace operator
\begin{equation}
O_{L,a_1\dots a_N}(x) =\text{Tr}\left[\phi_1^L(\psi_1)_{a_1}\cdots (\psi_1)_{a_N} \right](x)+ (\text{permutations})\,,
\end{equation}
in the double-scaling limit of $\gamma_i$-deformed $\mathcal{N}=4$ SYM theory \cite{Caetano:2016ydc}:
\begin{equation}
\gamma_1,\gamma_2  \to- i\infty \,,\,\,\gamma_3 \to i\infty\,,\, g\to 0\,|\,\,\,\, g^2 e^{i\gamma_1} = \xi_1^2\,,\,g^2 e^{i\gamma_2} = \xi_2^2\,.
\end{equation}
Thus, by the same arguments of the previous paragraphs, we can map the spectrum of the theory's scaling dimensions to the diagonalization of transfer matrices of the inhomogeneous spinning integrable chain with representation $(1,0,0)$, $\theta=-1/2$ and  $(3/2,0,1)$, $\theta=-3/4$.
\begin{figure}[H]
\includegraphics[scale=0.52]{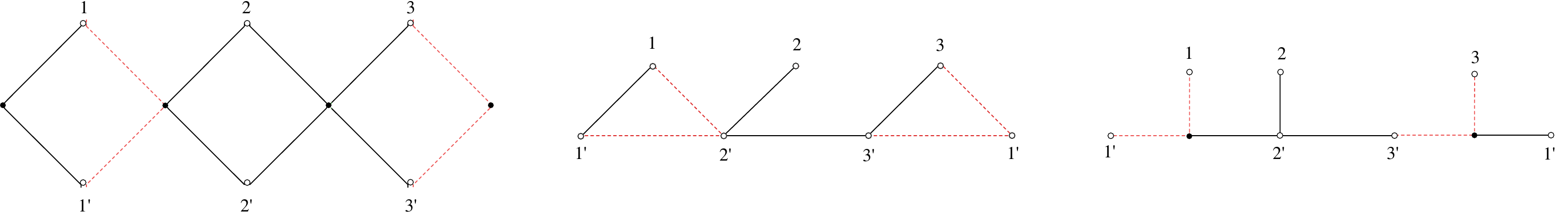}
\caption{\textbf{Left:} Integral kernel of the transfer matrix operator $t^{(a)}(u)$ for a spin chain of length $L=3$, for scalar auxiliary space $\ell_a=\dot{\ell}_a=0$ and inhomogeneous representation in physical spaces: scalar $(1,0,0)$ or  Weyl fermion $(3/2,0,1)$, with inhomogeinities $\theta=-\Delta/2$ according to \eqref{inhom_constr}. \textbf{Center:} reduction of the kernel at the point $u\to-1/2$. \textbf{Right:} by star-triangle duality the reduced kernel can be written as the graph building operator for a planar Feynman diagram wrapped onto a cylinder, whose topology mixes square-lattice and Yukawa hexagonal lattice.}
\end{figure}

\section{Outlooks}

The methods developed in this paper shall find application to multiple open problems. The generalization of star-triangle duality to any spinning propagator is a crucial step in order to mimic the $2d$ techniques of \cite{Derkachov:2001yn} and construct the transformation to separated
variables \cite{Sklyanin:1995bm} for the integrable model of section \ref{InFish}. Indeed, the attentive reader has certainly noticed a striking similarity between the integral kernels of Fig.\ref{t_pm_diag} and the eigenfunctions of the $B(u)$ operator in the $SL(2,\mathbb{C})$ chain.
The importance of constructing the separation of variables (SoV) transform in the $SO(1,5)$ chain is multiple: first the recent achievements of \cite{gromov2020determinant,Gromov:2019wmz,
Cavaglia:2019pow,Gromov:2016itr,Ryan:2020rfk,Ryan:2018fyo,
MailletNiccoliNewQSoVIntoruction,Maillet:2020ykb,Maillet:2019nsy,Derkachov:2018ewi} would be extended to the principal series representations - that appears in any operator product expansion in a CFT \cite{DiFrancesco1997,Tod:1977harm}. Moreover as a consequence of the considerations of section \ref{subsect:fish} the SoV would lead to an interesting example of direct derivation of a quantum spectral curve in the Fishnet and chiral conformal field theories \cite{Gurdogan:2015csr,Caetano:2016ydc,Kazakov_2019,Pittelli2019}.

Another captivating model to study by these new means is the open boundary version of the inhomogeneous spinning fishnet. In analogy with the computation of Basso-Dixon diagrams \cite{Derkachov2019,Derkachov2020,Derkachov:2020zvv}, a general inhomogeneous and spinning version of the fishnet lattice can be used to compute $4$-point correlators in the double-scaling (DS) limit of $\mathcal{N}=4$ SYM \cite{Gurdogan:2015csr,Kazakov_2018}, a richer theory than the bi-scalar fishnet that is a promising starting point to understand integrable structures in the undeformed theory. Similar considerations apply to higher-point correlators, where the technique of tessellations \cite{basso2015structure,Fleury:2016ykk,Fleury:2017eph,Eden:2016xvg,Basso_2019} may be repeated with a direct approach - the computation of Feynman integrals - in various DS limits of the SYM theory. This perspective is particularly appealing as it promises to shed light on the relation between the SoV and the computation of hexagon form factors.

Finally, the formulation of a fishnet theory in any even space-time dimension $d$ \cite{Kazakov:2018qez,BassoFerrando} leaves the question open about how to extend the star-triangle dualities, diagrams computation, separation of variables, from $SL(2,\mathbb{C})\simeq SO(1,3)$ and $SO(1,5)$ to the conformal spin chain $SO(1,d+1)$.

\section*{Ackowledgements}
We thank A.N.~Manashov and especially G.~Ferrando for careful reading and useful comments on the manuscript.
The work of S.D. was supported by the Russian Science Foundation project No 19-11-00131. Research at the Perimeter Institute is supported in part by the Government of Canada through NSERC and by the Province of Ontario through MRI. This work was additionally supported by a grant from the Simons Foundation (Simons Collaboration on the Nonperturbative Bootstrap).

\bibliographystyle{nb}
\bibliography{biblio_v2}

%bibliography generated by nb.bst v1.01 (C) 2003-2010 Niklas Beisert
\begin{thebibliography}{10}
\ifx\href\asklfhas\newcommand{\href}[2]{#2}\fi
\ifx\arxivref\asklfhas\newcommand{\arxivref}[2]{\href{http://arxiv.org/abs/#1}{#2}}\fi
\ifx\doiref\asklfhas\newcommand{\doiref}[2]{\href{http://dx.doi.org/#1}{#2}}\fi
\raggedright
\small
\parskip 0pt

\bibitem{Gurdogan:2015csr}
O.~Gurdogan and V.~Kazakov,
\textit{``{New integrable non-gauge 4D QFTs from strongly deformed planar N=4
  SYM}''},
\texttt{\arxivref{1512.06704}{arXiv:1512.06704}}.
%%CITATION = ARXIV:1512.06704;%%

\bibitem{Isaev_2003}
A.~Isaev,
\textit{``Multi-loop Feynman integrals and conformal quantum mechanics''},
\textsf{\doiref{10.1016/s0550-3213(03)00393-6}{Nuclear~Physics~B~662,~461–475~(2003)}},
\href{http://dx.doi.org/10.1016/S0550-3213(03)00393-6}{\texttt{http://dx.doi.org/10.1016/S0550-3213(03)00393-6}}.

\bibitem{Chicherin:2012yn}
D.~Chicherin, S.~Derkachov and A.~P.~Isaev,
\textit{``{Conformal group: R-matrix and star-triangle relation}''},
\textsf{\doiref{10.1007/JHEP04(2013)020}{JHEP~1304,~020~(2013)}},
\texttt{\arxivref{1206.4150}{arXiv:1206.4150}}.
%%CITATION = ARXIV:1206.4150;%%

\bibitem{Derkachov2020}
S.~Derkachov and E.~Olivucci,
\textit{``Exactly Solvable Magnet of Conformal Spins in Four Dimensions''},
\textsf{\doiref{10.1103/physrevlett.125.031603}{Physical~Review~Letters~125,~E.~Olivucci~(2020)}},
\href{http://dx.doi.org/10.1103/PhysRevLett.125.031603}{\texttt{http://dx.doi.org/10.1103/PhysRevLett.125.031603}}.

\bibitem{Derkachov:2020zvv}
S.~Derkachov and E.~Olivucci,
\textit{``{Exactly solvable single-trace four point correlators in
  $\chi$CFT$_4$}''},
\textsf{\doiref{10.1007/JHEP02(2021)146}{JHEP~2102,~146~(2021)}},
\texttt{\arxivref{2007.15049}{arXiv:2007.15049}}.

\bibitem{Isaev:2007uy}
A.~P.~Isaev,
\textit{``{Operator approach to analytical evaluation of Feynman diagrams}''},
\textsf{\doiref{10.1134/S1063778808050219}{Phys.~Atom.~Nucl.~71,~914~(2008)}},
\texttt{\arxivref{0709.0419}{arXiv:0709.0419}},
in: \textit{``{Proceedings, XII International Conference on Symmetry Methods in
  Physics: Yerevan, Armenia, July 3-8, 2006}''},
914-924p.
%%CITATION = ARXIV:0709.0419;%%

\bibitem{Vasiliev:1982dc}
A.~N.~Vasiliev, Y.~M.~Pismak and Y.~R.~Khonkonen,
\textit{``{1/N EXPANSION: CALCULATION OF THE EXPONENT ETA IN THE ORDER 1/N**3
  BY THE CONFORMAL BOOTSTRAP METHOD}''},
\textsf{\doiref{10.1007/BF01015292}{Theor.~Math.~Phys.~50,~127~(1982)}}.

\bibitem{Vasiliev:1981dg}
A.~N.~Vasiliev, Y.~M.~Pismak and Y.~R.~Khonkonen,
\textit{``{1/$N$ Expansion: Calculation of the Exponents $\eta$ and Nu in the
  Order 1/$N^2$ for Arbitrary Number of Dimensions}''},
\textsf{\doiref{10.1007/BF01019296}{Theor.~Math.~Phys.~47,~465~(1981)}}.

\bibitem{Kazakov:1983ns}
D.~I.~Kazakov,
\textit{``{Calculation of Feynman integrals by the method of
  ‘uniqueness'}''},
\textsf{\doiref{10.1007/BF01018044}{Theor.~Math.~Phys.~58,~223~(1984)}},
[Teor. Mat. Fiz.58,343(1984)].
%%CITATION = TMPHA,58,223;%%

\bibitem{Kazakov:1983pk}
D.~I.~Kazakov,
\textit{``{Multiloop Calculations: Method of Uniqueness and Functional
  Equations}''},
\textsf{\doiref{10.1007/BF01034829}{Theor.~Math.~Phys.~62,~84~(1985)}},
[Teor. Mat. Fiz.62,127(1984)].
%%CITATION = TMPHA,62,84;%%

\bibitem{Kazakov:1984km}
D.~I.~Kazakov,
\textit{``{The method of uniqueness, a new powerful technique for multiloop
  calculations}''},
\textsf{\doiref{10.1016/0370-2693(83)90816-X}{Phys.~Lett.~133B,~406~(1983)}}.
%%CITATION = PHLTA,133B,406;%%

\bibitem{Preti:2018vog}
M.~Preti,
\textit{``{STR: a Mathematica package for the method of uniqueness}''},
\texttt{\arxivref{1811.04935}{arXiv:1811.04935}}.
%%CITATION = ARXIV:1811.04935;%%

\bibitem{Preti_2020}
M.~Preti,
\textit{``The Game of Triangles''},
\textsf{\doiref{10.1088/1742-6596/1525/1/012015}{Journal~of~Physics:~Conference~Series~1525,~012015~(2020)}},
\href{http://dx.doi.org/10.1088/1742-6596/1525/1/012015}{\texttt{http://dx.doi.org/10.1088/1742-6596/1525/1/012015}}.

\bibitem{Tod:1977harm}
V.~K.~Dobrev et~al.,
\textit{``Harmonic Analysis on the n-Dimensional Lorentz Group and Its
  Application to Conformal Quantum Field Theory''},
\textsf{\doiref{10.1007/BFb0009678}{Lect.Notes~Phys.~63~12,~059~(1977)}}.

\bibitem{Derkachov:2001yn}
S.~E.~Derkachov, G.~P.~Korchemsky and A.~N.~Manashov,
\textit{``{Noncompact Heisenberg spin magnets from high-energy QCD: 1. Baxter Q
  operator and separation of variables}''},
\textsf{\doiref{10.1016/S0550-3213(01)00457-6}{Nucl.~Phys.~B617,~375~(2001)}},
\texttt{\arxivref{0107193}{arXiv:0107193}}.
%%CITATION = HEP-TH/0107193;%%

\bibitem{Caetano:2016ydc}
J.~Caetano, O.~Gurdogan and V.~Kazakov,
\textit{``{Chiral limit of N = 4 SYM and ABJM and integrable Feynman
  graphs}''},
\texttt{\arxivref{1612.05895}{arXiv:1612.05895}}.
%%CITATION = ARXIV:1612.05895;%%

\bibitem{Kazakov_2018}
V.~Kazakov,
\textit{``Quantum Spectral Curve of gamma-Twisted N= 4 SYM Theory and Fishnet
  CFT''},
\textsf{\doiref{10.1142/s0129055x1840010x}{Reviews~in~Mathematical~Physics~30,~1840010~(2018)}},
\href{http://dx.doi.org/10.1142/S0129055X1840010X}{\texttt{http://dx.doi.org/10.1142/S0129055X1840010X}}.

\bibitem{Kazakov_2019}
V.~Kazakov, E.~Olivucci and M.~Preti,
\textit{``Generalized fishnets and exact four-point correlators in chiral
  CFT4''},
\textsf{\doiref{10.1007/jhep06(2019)078}{Journal~of~High~Energy~Physics~2019,~M.~Preti~(2019)}},
\href{http://dx.doi.org/10.1007/JHEP06(2019)078}{\texttt{http://dx.doi.org/10.1007/JHEP06(2019)078}}.

\bibitem{Pittelli2019}
A.~Pittelli and M.~Preti,
\textit{``Integrable fishnet from gamma-deformed N=2 quivers''},
\textsf{\doiref{10.1016/j.physletb.2019.134971}{Physics~Letters~B~798,~134971~(2019)}},
\href{http://dx.doi.org/10.1016/j.physletb.2019.134971}{\texttt{http://dx.doi.org/10.1016/j.physletb.2019.134971}}.

\bibitem{Lipatov:1993qn}
L.~N.~Lipatov,
\textit{``{High-energy asymptotics of multicolor QCD and two-dimensional
  conformal field theories}''},
edited by J.~Tran Thanh~Van,
\textsf{\doiref{10.1016/0370-2693(93)90951-D}{Phys.~Lett.~B~309,~394~(1993)}}.

\bibitem{Lipatov:1993yb}
L.~N.~Lipatov,
\textit{``{Asymptotic behavior of multicolor QCD at high energies in connection
  with exactly solvable spin models}''},
\textsf{JETP~Lett.~59,~596~(1994)},
\texttt{\arxivref{hep-th/9311037}{hep-th/9311037}}.

\bibitem{Faddeev:1994zg}
L.~D.~Faddeev and G.~P.~Korchemsky,
\textit{``{High-energy QCD as a completely integrable model}''},
\textsf{\doiref{10.1016/0370-2693(94)01363-H}{Phys.~Lett.~B~342,~311~(1995)}},
\texttt{\arxivref{hep-th/9404173}{hep-th/9404173}}.

\bibitem{Kazakov:2018qez}
V.~Kazakov and E.~Olivucci,
\textit{``{Biscalar Integrable Conformal Field Theories in Any Dimension}''},
\textsf{\doiref{10.1103/PhysRevLett.121.131601}{Phys.~Rev.~Lett.~121,~131601~(2018)}},
\texttt{\arxivref{1801.09844}{arXiv:1801.09844}}.
%%CITATION = ARXIV:1801.09844;%%

\bibitem{Gromov:2017cja}
N.~Gromov, V.~Kazakov, G.~Korchemsky, S.~Negro and G.~Sizov,
\textit{``{Integrability of Conformal Fishnet Theory}''},
\textsf{\doiref{10.1007/JHEP01(2018)095}{JHEP~1801,~095~(2018)}},
\texttt{\arxivref{1706.04167}{arXiv:1706.04167}}.

\bibitem{Derkachov2019}
S.~Derkachov, V.~Kazakov and E.~Olivucci,
\textit{``Basso-Dixon correlators in two-dimensional fishnet CFT''},
\textsf{\doiref{10.1007/jhep04(2019)032}{Journal~of~High~Energy~Physics~2019,~E.~Olivucci~(2019)}},
\href{http://dx.doi.org/10.1007/JHEP04(2019)032}{\texttt{http://dx.doi.org/10.1007/JHEP04(2019)032}}.

\bibitem{BassoFerrando}
B.~Basso, G.~Ferrando, V.~Kazakov and D.-l.~Zhong,
\textit{``Thermodynamic Bethe Ansatz for Biscalar Conformal Field Theories in
  any Dimension''},
\textsf{\doiref{10.1103/physrevlett.125.091601}{Physical~Review~Letters~125,~D.~(2020)}},
\href{http://dx.doi.org/10.1103/PhysRevLett.125.091601}{\texttt{http://dx.doi.org/10.1103/PhysRevLett.125.091601}}.

\bibitem{Gelfand:105396}
I.~M.~Gelfand and G.~E.~Shilov,
\textit{``{Generalized functions, Vol I}''},
Academic Press (1964),
New York, NY,
Trans. from the Russian, Moscow, 1958.

\bibitem{Sotkov:1976xe}
G.~M.~Sotkov and R.~P.~Zaikov,
\textit{``{Conformal Invariant Two Point and Three Point Functions for Fields
  with Arbitrary Spin}''},
\textsf{\doiref{10.1016/0034-4877(77)90033-7}{Rept.~Math.~Phys.~12,~375~(1977)}}.

\bibitem{Kulish:1981gi}
P.~P.~Kulish, N.~Y.~Reshetikhin and E.~K.~Sklyanin,
\textit{``{Yang-Baxter Equation and Representation Theory. 1.}''},
\textsf{\doiref{10.1007/BF02285311}{Lett.~Math.~Phys.~5,~393~(1981)}}.

\bibitem{Kulish:1981bi}
P.~P.~Kulish and E.~K.~Sklyanin,
\textit{``{QUANTUM SPECTRAL TRANSFORM METHOD. RECENT DEVELOPMENTS}''},
edited by J.~Hietarinta and C.~Montonen,
\textsf{Lect.~Notes~Phys.~151,~61~(1982)}.

\bibitem{Faddeev:1996iy}
L.~D.~Faddeev,
\textit{``{How algebraic Bethe ansatz works for integrable model}''},
\texttt{\arxivref{9605187}{arXiv:9605187}},
in: \textit{``{Connes, A. (ed.) et al., Quantum symmetries/ Symétries
  quantiques. Proceedings of the Les Houches summer school, Session LXIV, Les
  Houches, France, August 1 - September 8, 1995. Amsterdam:
  North-Holland.(1998)}''},
149-219p.
%%CITATION = HEP-TH/9605187;%%

\bibitem{Kulish:1980ii}
P.~P.~Kulish and E.~K.~Sklyanin,
\textit{``{On the solution of the Yang-Baxter equation}''},
\textsf{\doiref{10.1007/BF01091463}{J.~Sov.~Math.~19,~1596~(1982)}}.

\bibitem{Tarasov:1983cj}
V.~O.~Tarasov, L.~A.~Takhtajan and L.~D.~Faddeev,
\textit{``{Local Hamiltonians for integrable quantum models on a lattice}''},
\textsf{\doiref{10.1007/BF01018648}{Theor.~Math.~Phys.~57,~1059~(1983)}}.

\bibitem{Gromov:2018hut}
N.~Gromov, V.~Kazakov and G.~Korchemsky,
\textit{``{Exact Correlation Functions in Conformal Fishnet Theory}''},
\texttt{\arxivref{1808.02688}{arXiv:1808.02688}}.
%%CITATION = ARXIV:1808.02688;%%

\bibitem{Gromov2020}
N.~Gromov and A.~Sever,
\textit{``The holographic dual of strongly gamma-deformed $\mathcal{N}$ = 4 SYM
  theory: derivation, generalization, integrability and discrete
  reparametrization symmetry''},
\textsf{\doiref{10.1007/jhep02(2020)035}{Journal~of~High~Energy~Physics~2020,~A.~Sever~(2020)}},
\href{http://dx.doi.org/10.1007/JHEP02(2020)035}{\texttt{http://dx.doi.org/10.1007/JHEP02(2020)035}}.

\bibitem{Basso:2017jwq}
B.~Basso and L.~J.~Dixon,
\textit{``{Gluing Ladder Feynman Diagrams into Fishnets}''},
\textsf{\doiref{10.1103/PhysRevLett.119.071601}{Phys.~Rev.~Lett.~119,~071601~(2017)}},
\texttt{\arxivref{1705.03545}{arXiv:1705.03545}}.
%%CITATION = ARXIV:1705.03545;%%

\bibitem{Chicherin:2017frs}
D.~Chicherin, V.~Kazakov, F.~Loebbert, D.~Mueller and D.-l.~Zhong,
\textit{``{Yangian Symmetry for Fishnet Feynman Graphs}''},
\textsf{\doiref{10.1103/PhysRevD.96.121901}{Phys.~Rev.~D96,~121901~(2017)}},
\texttt{\arxivref{1708.00007}{arXiv:1708.00007}}.
%%CITATION = ARXIV:1708.00007;%%

\bibitem{Chicherin:2017cns}
D.~Chicherin, V.~Kazakov, F.~Loebbert, D.~Mueller and D.-l.~Zhong,
\textit{``{Yangian Symmetry for Bi-Scalar Loop Amplitudes}''},
\texttt{\arxivref{1704.01967}{arXiv:1704.01967}}.
%%CITATION = ARXIV:1704.01967;%%

\bibitem{Sklyanin:1995bm}
E.~K.~Sklyanin,
\textit{``{Separation of variables - new trends}''},
\textsf{\doiref{10.1143/PTPS.118.35}{Prog.~Theor.~Phys.~Suppl.~118,~35~(1995)}},
\texttt{\arxivref{9504001}{arXiv:9504001}}.
%%CITATION = SOLV-INT/9504001;%%

\bibitem{gromov2020determinant}
N.~Gromov, F.~Levkovich-Maslyuk and P.~Ryan,
\textit{``Determinant Form of Correlators in High Rank Integrable Spin Chains
  via Separation of Variables''},
\texttt{\arxivref{2011.08229}{arXiv:2011.08229}}.

\bibitem{Gromov:2019wmz}
N.~Gromov, F.~Levkovich-Maslyuk, P.~Ryan and D.~Volin,
\textit{``{Dual Separated Variables and Scalar Products}''},
\textsf{\doiref{10.1016/j.physletb.2020.135494}{Phys.~Lett.~B~806,~135494~(2020)}},
\texttt{\arxivref{1910.13442}{arXiv:1910.13442}}.

\bibitem{Cavaglia:2019pow}
A.~Cavagli\`a, N.~Gromov and F.~Levkovich-Maslyuk,
\textit{``{Separation of variables and scalar products at any rank}''},
\textsf{\doiref{10.1007/JHEP09(2019)052}{JHEP~1909,~052~(2019)}},
\texttt{\arxivref{1907.03788}{arXiv:1907.03788}}.

\bibitem{Gromov:2016itr}
N.~Gromov, F.~Levkovich-Maslyuk and G.~Sizov,
\textit{``{New Construction of Eigenstates and Separation of Variables for
  SU(N) Quantum Spin Chains}''},
\textsf{\doiref{10.1007/JHEP09(2017)111}{JHEP~1709,~111~(2017)}},
\texttt{\arxivref{1610.08032}{arXiv:1610.08032}}.

\bibitem{Ryan:2020rfk}
P.~Ryan and D.~Volin,
\textit{``{Separation of variables for rational gl(n) spin chains in any
  compact representation, via fusion, embedding morphism and Backlund flow}''},
\texttt{\arxivref{2002.12341}{arXiv:2002.12341}}.

\bibitem{Ryan:2018fyo}
P.~Ryan and D.~Volin,
\textit{``{Separated variables and wave functions for rational gl(N) spin
  chains in the companion twist frame}''},
\textsf{\doiref{10.1063/1.5085387}{J.~Math.~Phys.~60,~032701~(2019)}},
\texttt{\arxivref{1810.10996}{arXiv:1810.10996}}.

\bibitem{MailletNiccoliNewQSoVIntoruction}
J.~M.~Maillet and G.~Niccoli,
\textit{``{"On quantum separation of variables."}''},
\textsf{J.~Math.~Phys.~\bf 59,~091417~(2018)}.

\bibitem{Maillet:2020ykb}
J.~M.~Maillet, G.~Niccoli and L.~Vignoli,
\textit{``{On Scalar Products in Higher Rank Quantum Separation of
  Variables}''},
\textsf{\doiref{10.21468/SciPostPhys.9.6.086}{SciPost~Phys.~9,~086~(2020)}},
\texttt{\arxivref{2003.04281}{arXiv:2003.04281}}.

\bibitem{Maillet:2019nsy}
J.~M.~Maillet and G.~Niccoli,
\textit{``{On quantum separation of variables beyond fundamental
  representations}''},
\texttt{\arxivref{1903.06618}{arXiv:1903.06618}}.

\bibitem{Derkachov:2018ewi}
S.~E.~Derkachov and P.~A.~Valinevich,
\textit{``{Separation of variables for the quantum $SL(3,\mathbb C)$ spin
  magnet: eigenfunctions of Sklyanin $B$-operator}''},
\textsf{\doiref{10.1007/s10958-019-04505-5}{J.~Math.~Sci.~242,~658~(2019)}},
\texttt{\arxivref{1807.00302}{arXiv:1807.00302}}.

\bibitem{DiFrancesco1997}
P.~Di~Francesco, P.~Mathieu and D.~Senechal,
\textit{``{Conformal Field Theory}''},
Springer-Verlag (1997),
New York.

\bibitem{basso2015structure}
B.~Basso, S.~Komatsu and P.~Vieira,
\textit{``Structure Constants and Integrable Bootstrap in Planar N=4 SYM
  Theory''},
\texttt{\arxivref{1505.06745}{arXiv:1505.06745}}.

\bibitem{Fleury:2016ykk}
T.~Fleury and S.~Komatsu,
\textit{``{Hexagonalization of Correlation Functions}''},
\textsf{\doiref{10.1007/JHEP01(2017)130}{JHEP~1701,~130~(2017)}},
\texttt{\arxivref{1611.05577}{arXiv:1611.05577}}.
%%CITATION = ARXIV:1611.05577;%%

\bibitem{Fleury:2017eph}
T.~Fleury and S.~Komatsu,
\textit{``{Hexagonalization of Correlation Functions II: Two-Particle
  Contributions}''},
\textsf{\doiref{10.1007/JHEP02(2018)177}{JHEP~1802,~177~(2018)}},
\texttt{\arxivref{1711.05327}{arXiv:1711.05327}}.
%%CITATION = ARXIV:1711.05327;%%

\bibitem{Eden:2016xvg}
B.~Eden and A.~Sfondrini,
\textit{``{Tessellating cushions: four-point functions in $\mathcal{N} $ = 4
  SYM}''},
\textsf{\doiref{10.1007/JHEP10(2017)098}{JHEP~1710,~098~(2017)}},
\texttt{\arxivref{1611.05436}{arXiv:1611.05436}}.
%%CITATION = ARXIV:1611.05436;%%

\bibitem{Basso_2019}
B.~Basso, J.~Caetano and T.~Fleury,
\textit{``Hexagons and correlators in the fishnet theory''},
\textsf{\doiref{10.1007/jhep11(2019)172}{Journal~of~High~Energy~Physics~2019,~T.~Fleury~(2019)}},
\href{http://dx.doi.org/10.1007/JHEP11(2019)172}{\texttt{http://dx.doi.org/10.1007/JHEP11(2019)172}}.

\end{thebibliography}

\end{document}